\documentclass[12pt]{article}
\usepackage{physics}
\usepackage{mathtools}
\usepackage{slashed}
\usepackage{graphicx,xcolor}
\usepackage{authblk}
\usepackage{color}

\usepackage[
 	backend=biber,
 	style=numeric-comp,
	sorting=none
]{biblatex}
\usepackage{hyperref}
\hypersetup{
 colorlinks=true,
 linkcolor=blue,
 citecolor=violet,
 filecolor=magenta,
 urlcolor=cyan
}
\hypersetup{linktocpage}
\usepackage{orcidlink}

\usepackage[T1]{fontenc}
\usepackage{amsmath}
\usepackage{amssymb}
\usepackage{slashed}
\usepackage{color}
\usepackage{setspace}

\begin{document}

\title{\bf Thermal corrections to dark matter annihilation with real
photon emission/absorption}


\author[1,2]{Prabhat Butola\footnote{prabhatb@imsc.res.in,
\orcidlink{0009-0003-2824-455X}}}

\author[1,2]{D. Indumathi\footnote{indu@imsc.res.in,
\orcidlink{0000-0001-6685-4760}}}

\author[3]{Pritam Sen\footnote{pritam.sen@tifr.res.in,
\orcidlink{0000-0003-0751-5560}}}


\affil[1]{Homi Bhabha National Institute, Mumbai, India}

\affil[2]{Institute of Mathematical Sciences, Chennai, India}

\affil[3]{Department of Theoretical Physics, Tata Institute of
Fundamental Research, Mumbai, India}

\maketitle

\abstract{%
The dark matter relic density is being increasingly precisely measured.
This relic density is theoretically determined by a Boltzmann equation
which computes the dark matter distribution according to the (thermally
averaged) cross section for annihilation/production of dark matter
within a given model. We present here the complete higher order thermal
corrections, calculated using real time thermal field theory, to the
cross section for the annihilation of dark matter into Standard Model
fermions via charged scalars: $\chi \chi \to f \overline{f}$ and $\chi
\chi \to f \overline{f} (\gamma)$. The latter process includes real photon
emission into, and absorption from, the heat bath at temperature $T$. We
use the Grammer and Yennie technique to separate the soft infra-red
divergences, which greatly simplifies the calculation. We show explicitly
the cancellation of both soft and collinear divergences between real and
virtual contributions at next-to-leading order (NLO) in the thermal field
theory and remark on the non-trivial nature of the collinear divergences
when the thermal contribution from fermions is considered. We present
the leading thermal contributions to order ${\cal{O}}(\alpha T^2)$
for the case when the dark matter particle is a Majorana or Dirac type
fermion. In the former case, both the leading order (LO) and NLO cross
sections are helicity suppressed, while neither is suppressed in the
Dirac case. It is interesting that the ratio of NLO to LO cross sections
is the same for both Majorana and Dirac type dark matter.}

\vspace{0.5cm}


\section{Introduction}
\label{sec:intro}

Thermal corrections to dark matter (DM) annihilation cross sections
are of interest due to the precision with which the dark matter
relic density has been measured, and given that temperature effects
are important in the early Universe. While evidence for DM arises from
several astrophysical and cosmological observations \cite{Planck:2018vyg,
Circiello:2024gpq, ManceraPina:2024ybj, Bechtol:2022koa, Amendola:2016saw,
Baudis:2015mpa}, its properties are not as yet fully determined
since they have not been directly detected; see, for instance,
Ref.~\cite{ParticleDataGroup:2024cfk}, for a review. However, the
existence of DM is considered to be well-established, with a precisely
measured relic density (its present abundance in the Universe),
$\Omega_c h^2 = 0.1200 \pm 0.0012$ \cite{Planck:2018vyg}, where $h$
is the reduced Hubble constant, $h = H_0/100$, and the subscript $c$
refers to cold dark matter.

There exist many models of DM with differing properties such as such as
weakly interacting massive particles, axion-like particles, neutralino
dark matter, sterile neutrinos, particles in extra dimensional theories,
primordial black holes, etc.; see Refs.~\cite{Kolb:1990vq, Griest:1990kh,
Bertone:2004pz, Bertone:2010zza, Bertone:2016nfn, Profumo:2017hqp,
Bertone:2018krk, Profumo:2019ujg, Feng:2022rxt} for reviews. There
are many scenarios like asymmetric dark matter, etc., with freeze-out
and freeze-in being two popular mechanisms. In the former case, the DM
is considered to be in thermal, chemical and kinetic equilibrium with
standard model (SM) particles until its interaction rate falls below the
Hubble expansion rate\footnote{There are several applications of the
so-called freeze-out scenario; Refs.~\cite{Cowsik:1972gh, Lee:1977ua}
discussed this possibility in the context of neutrinos.}, when the
DM freezes out, giving the present relic density. In the freeze-in
scenario, the coupling of DM to SM particles is small so that the
DM is never in equilibrium with the SM particles; the amount of dark
matter in the Universe then keeps on increasing until it freezes-in to
the present relic density \cite{Hall:2009bx, Bernal:2017kxu}. In both
cases, the theoretical quantity of interest is the collision term, which
determines how the relic density evolves using the Boltzmann equation
\cite{Gondolo:1990dk}. The collision term is simply related to the cross
section for dark matter annihilation/production.

The present-day experimental determination of the dark matter relic
density is now so precise that these model-based cross sections need to be
calculated to next-to-leading order (NLO); for example, the cross sections
in various models have been calculated in Refs.~\cite{Drees:2013er,
Harz:2014tma, Klasen:2016qux, Ala-Mattinen:2019mpa, Beneke:2019qaa,
Beneke:2020vff, Baumgart:2022vwr}. Additionally, {\em thermal
corrections} to these cross sections (or decay rates) can become
important in the early Universe; see Refs.~\cite{Czarnecki:2011mr,
Biondini:2013xua, Biondini:2015gyw, Biondini:2023zcz,
Beneke:2014gla, Binder:2021otw,Park:2025iog}, in Quark Gluon Plasma
\cite{Mustafa:2025uad}. Some of the processes have also been resummed
\cite{Jackson:2023zkl,Biondini:2025ihi, Bittar:2025lcr}. Other work
\cite{Laine:2022ner} has also discussed in detail the limitations of the
Boltzmann equation. In particular, Beneke et al.~\cite{Beneke:2014gla}
computed the thermal NLO corrections to the dark matter annihilation
cross section in a generic DM model where Majorana dark matter particles
interact with SM fermions through charged scalars where they showed
that the leading $s$-wave (non-relativistic) contribution to the thermal
cross section for the dark matter annihilation to SM fermion pairs via
scalars, {\em viz.}, $\chi \chi \to f \overline{f}$, of order ${\cal
O}(\alpha T^2)$, vanishes in the massless fermion limit, where $T$
is the temperature of the heat bath in which the interaction occurs.

In our earlier work~\cite{Butola:2024oia}, the higher order
$({\cal{O}}(\alpha)$) finite thermal contribution at order ${\cal
O}(T^2)$ was independently calculated for {\em virtual} corrections for
the same process. It was shown that the contribution was suppressed
due to helicity suppression, the same as with the leading order (LO)
cross section. In this paper, we calculate the ${\cal{O}}(\alpha)$
thermal cross section for the process, $\chi \chi \to f \overline{f}
(\gamma)$, that is, the {\em real} photon thermal corrections to the
annihilation cross section. Here the notation $(\gamma)$ indicates that
the real photon may be both emitted into, and absorbed by, the heat
bath. Although, in principle, there is now a three-body final state with
the possibility to lift the helicity suppression, in contrast to the LO
and the virtual NLO thermal corrections, we find that this process is
also helicity suppressed. In contrast, annihilation processes involving
Dirac (not Majorana) dark matter particles remain helicity un-suppressed
\cite{Hannestad:1999fj} at both the LO and NLO and can significantly
alter the collision term in the Boltzmann equation and hence alter the
dark matter relic density calculations.

In the earlier paper \cite{Butola:2024oia}, we had directly calculated
the {\em finite} part of the thermal virtual cross section using the
technique of Grammer and Yennie (GY) \cite{Grammer:1973db}, which
collects the infra-red (IR) finite terms into the so-called $G$-photon
contribution, while isolating the IR divergences into the so-called
$K$-photon terms. This technique was used \cite{Sen:2020oix, Sen:2018ybx}
to prove for the first time, the all-orders cancellation of the IR
divergences between the virtual and real contributions, order by order,
to all orders in perturbation theory. In this paper, we have calculated
the complete real thermal contribution to the dark matter annihilation
process; hence we also demonstrate here explicitly the cancellation at NLO
of IR divergences between the real and virtual contributions. However,
while doing so, we found that the GY technique only separates the
{\em soft} IR divergences, not the {\em collinear} ones. We show here
explicitly the cancellation of the collinear divergences between the real
and virtual contributions, diagram by diagram, at NLO. The structure of
these collinear divergences turns out to be non-trivial and interesting.

In the next Section, we review in brief the generic dark matter model that
was used in our earlier work \cite{Butola:2024oia} for calculating the
NLO thermal contribution to the virtual annihilation cross section. For
completeness, we also list the LO and virtual NLO cross sections that were
obtained there. Since the details of the $K$ and $G$ photon separation
become important in the handling of the IR divergences, we present the
key elements of the Grammer and Yennie technique as applicable to thermal
field theory in Section~\ref{sec:GY} for later use. (Appendix~\ref{sec:tft}
contains some relevant details and Feynman rules for calculations in
the real-time formulation of thermal field theory.)

In Section~\ref{sec:real} we present the results for the
real photon thermal contribution to the dark matter annihilation
cross section $\chi \chi \to f \overline{f} (\gamma)$ where the
$(\gamma)$ indicates that the photon may be emitted into, or absorbed
from, the heat bath at temperature $T$. We use the FeynCalc 10.0.0
\cite{Shtabovenko:2023idz, Mertig:1990an} software with Mathematica
13.1 \cite{Mathematica} for the calculations. We present explicitly
here, the cancellation of the soft IR and collinear divergences between
the real and virtual contributions. Since the virtual contributions are
necessary for demonstrating this cancellation, we present some highlights
of this calculation, including a discussion of self energy diagrams, in
Appendix~\ref{sec:virtual} for completeness. We present our conclusions
with discussions in Section~\ref{sec:concl}.

\section{The model, and the virtual NLO thermal corrections, in brief}
\label{sec:modelv}

We briefly review earlier work \cite{Butola:2024oia} where the virtual
NLO thermal contribution to the dark matter annihilation cross section
$\chi \chi \to f \overline{f}$ was calculated using a generic Lagrangian
of the form \cite{DiFranzo:2013vra, Beneke:2014gla}
\begin{align}
{\cal{L}} & = -\frac{1}{4} F_{\mu\nu} F^{\mu\nu} +
\overline{{f}} \left( i \slashed{D} - m_f \right) {{f}} +
\frac{1}{2} \overline{\chi} \left( i \slashed{\partial} - m_\chi \right)
\chi \nonumber \\
 & \qquad + \left(D^\mu \phi \right)^\dagger \left(D_\mu \phi \right) -
 m_\phi^2 \phi^\dagger \phi + \left( \lambda \overline{\chi} P_L {{f}}^-
 \phi^+ + {\rm h.c.} \right)~.
\label{eq:L}
\end{align}
Here the dark matter candidate is an $SU(2) \times U(1)$ singlet
Majorana fermion $\chi$ which interacts with SM doublet fermions, $f =
(f^0,f^-)^T$, via scalar partners $\phi = (\phi^+, \phi^0)^T$ through
a Yukawa interaction. The calculations are performed after electro-weak
symmetry breaking, so that the $W$ and $Z$ are heavy and their
contributions to the higher order corrections can be safely ignored.

The NLO thermal corrections to the annihilation of dark matter via
$\chi \chi \to f \overline{f}$ in this model were first studied in
Ref.~\cite{Beneke:2014gla} where the cancellation of the infra-red
divergences was explicitly demonstrated and the finite remainder
calculated to leading order in the temperature, ${\cal O}(T^2)$. This
was then extended in Refs.~\cite{Sen:2020oix, Sen:2018ybx} to a proof of
the cancellation of infra-red divergences to all orders in such thermal
field theories, using a generalised Grammer and Yennie (GY) technique
although the finite remainder was not calculated. The advantage of the
GY approach, details of which we present in the next section, is the
automatic cancellation of the soft infra-red (IR) divergences so that
we can directly compute the finite remainder. This approach was used to
calculate the explicitly IR finite remainder in the NLO thermal virtual
correction to the dark matter annihilation process, $\chi \chi \to f
\overline{f}$, in our earlier work~\cite{Butola:2024oia}.

\begin{figure}[htp]
\begin{center}
\includegraphics[width=0.3\textwidth]{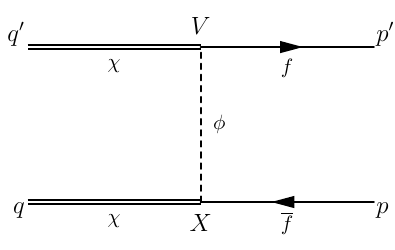}
\hspace{1cm}
\includegraphics[width=0.3\textwidth]{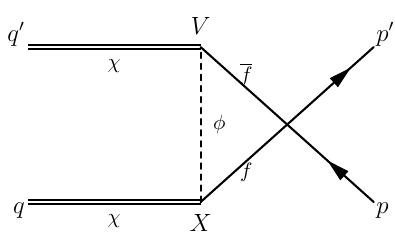}
\caption{\small \em The $t$-channel and $u$-channel dark matter
annihilation processes at leading order (LO).}
\label{fig:lo}
\end{center}
\end{figure}

The LO annihilation contribution is shown in Fig.~\ref{fig:lo}; note
that since the dark matter particle is assumed to be a Majorana fermion,
both $t$ and $u$ channel diagrams contribute.

We work in the centre-of-momentum frame where the dark matter particles
with momenta $q'$ and $q$ collide back-to-back. In the leading order,
$P$ and $Q$ are the magnitude of the three-momenta of the fermion (or
anti-fermion) and dark matter particles respectively, $s$ is the usual
Mandelstam variable, $s = 4 H^2$, $H^2 = Q^2 + m_\chi^2 = P^2 + m_f^2$,
and $m_i, i = \phi, f, \chi$, are the masses of the scalar, fermion,
and dark matter particles respectively.

We perform the calculation in the limit in which the scalar is much heavier
than the other particles, {\em viz.}, $m_\phi^2 \gg m_\chi^2 \gtrsim
m_f^2$. In this limit, the momentum of the scalar, (which is $l \equiv
q - p$ in the $t$-channel and $l' \equiv q' -p$ in the $u$ channel)
is such that $l^2 \sim l'^2 \ll m_\phi^2$ (where we have implicitly
assumed that $\sqrt{s} \ll m_\phi$) so that the scalar propagator can
be approximated by $i D_{l_\phi} = i/(l^2-m_\phi^2) \to -i/m_\phi^2$;
then the LO cross section within this approximation can be expressed as
\cite{Butola:2024oia} 
\begin{align}
\sigma_{LO}^{\rm Heavy~Scalar} & =
\frac{1}{32\pi s} \frac{P}{Q} \int \!{\rm d}\!\cos\theta~
\left \vert \rule{0pt}{12pt} {\cal{M}}_{LO}^t - {\cal{M}}_{LO}^u
\right \vert^2~, \nonumber \\
 & =
\frac{1}{12\pi s} \frac{P}{Q}
\frac{\lambda^4}{m_\phi^4}
 \left[8 H^2( H^2 - m_\chi^2) +m_f^2 (5 m_\chi^2 - 2 H^2)\right]~,
\label{eq:sigma_LO}
\end{align}
where ${\cal{M}}_{LO}^t$ and ${\cal{M}}_{LO}^u$ are the matrix elements
corresponding to the LO $t$-channel and $u$-channel diagrams shown in
Fig.~\ref{fig:lo}. The first term in the square brackets is proportional
to $H^2 - m_\chi^2 \equiv Q^2$ and so vanishes in the non-relativistic
limit; the second is proportional to $m_f^2$ and is therefore helicity
suppressed because of the nature of the coupling (see Eq.~\ref{eq:L}). In
contrast, the purely $t$-channel contribution (relevant for a Dirac-type
dark matter particle) is not helicity suppressed; we have
\begin{align}
\sigma_{LO}^{\rm Dirac, heavy~scalar} & =
\frac{1}{12\pi s}
\frac{P}{Q}
\frac{\lambda^4}{m_\phi^4}
 \left[H^2( 4 H^2 - m_\chi^2) - m_f^2 (H^2 - m_\chi^2)\right]~.
\label{eq:sigma_LODirac}
\end{align}
Writing $Q \approx m_\chi v$, and replacing the CM velocity $v$ by the
relative velocity between the two dark matter particles, $v_{rel} =
2 v$, the Majorana LO cross section can be written as
\begin{align}
\sigma_{LO} v_{rel} 
 & \overset{Heavy~Scalar}{\xrightarrow{~~~~~~~~~~~~}}
 \frac{\lambda^4}{48 \pi} \left[\frac{1} {m_\phi^4} \right] 
 \left[6 m_f^2 + v^2\left(16 m_\chi^2 - 7 m_f^2 \right)\right]
 \left[1-\frac{m_f^2}{2 m_\chi^2} \right]~,
\label{eq:sigma_LOv}
\end{align}
where we have retained terms upto order ${\cal{O}}(v^2)$.

The thermal average of this quantity is the collision term of interest
in the Boltzmann equation to compute the dark matter relic density:
to leading order we have\footnote{This is the thermally averaged
centre-of-momentum cross section; a further multiplying factor is required
to convert it to the actual lab-frame collision term; however, here we are
primarily interested in the powers of temperature of the various terms,
or rather, powers of $x^{-1} = T/m_\chi$.}
\begin{align}
\langle\sigma_{LO} v_{rel} \rangle_T
 & \overset{Heavy~Scalar}{\xrightarrow{~~~~~~~~~~~~}}
 \frac{\lambda^4}{48 \pi} \left[\frac{1} {m_\phi^4} \right] 
 \left[6 m_f^2 + \frac{3 T}{2 m_\chi}
 \left(16 m_\chi^2 - 7 m_f^2 \right)\right]
 \left[1-\frac{m_f^2}{2 m_\chi^2} \right]~,
\label{eq:sigma_coll}
\end{align}
where we have used the non-relativistic Maxwell-Boltzmann distribution
function \cite{Gondolo:1990dk} for the dark matter particles so that
$\langle v^2 \rangle_T = 3T/(2m_\chi)$.

\subsection{The thermal NLO virtual contribution to the annihilation
cross section}

\begin{figure}[hbp]
\begin{center}
\includegraphics[width=0.25\textwidth]{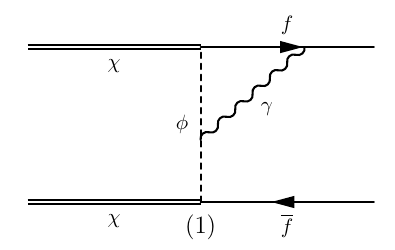}
\includegraphics[width=0.25\textwidth]{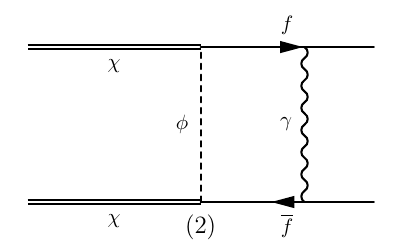}
\includegraphics[width=0.25\textwidth]{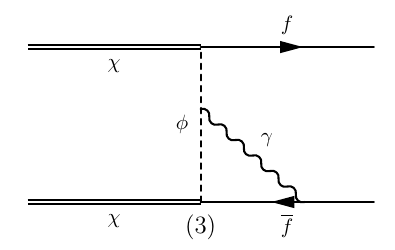}
\includegraphics[width=0.25\textwidth]{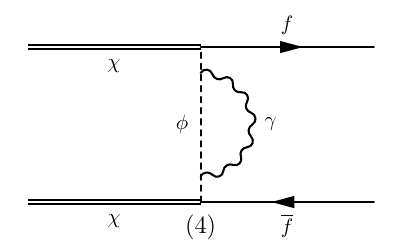}
\includegraphics[width=0.25\textwidth]{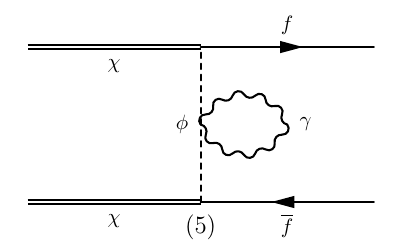}
\caption{\small \em The $t$-channel virtual photon corrections to the
dark matter annihilation process at NLO. Diagrams are labelled from
1--5. Analogous contributions from the $u$-channel diagrams exist.}
\label{fig:nlo}
\end{center}
\end{figure}

The NLO virtual corrections arise from the combined contributions
of ${\cal{M}}_{LO}^{t,u}$, the LO diagrams in Fig.~\ref{fig:lo} and
the contributions ${\cal{M}}_{NLO}^{1\hbox{--}5}$ from the set of
NLO diagrams labelled 1--5 in Fig.~\ref{fig:nlo} with virtual photon
of momentum $k$.

We use the Feynman rules for thermal field theory given in
Appendix~\ref{sec:tft}. In the real-time formulation of thermal field
theory that we use here, there is a field-doubling with type-1 and type-2
thermal fields, so that the propagator is a $2\times 2$ matrix. The
propagator can be expressed as a sum of $T$-independent and $T$-dependent
parts; the $T$-dependent part allows for a transition between the two
types of fields. In particular, it gives rise to the corresponding
thermal contribution in the case of virtual photon NLO corrections. All
external particles are of type-1 alone; see Appendix~\ref{sec:tft}
for more details.

Note that the so-called leading order diagrams in Fig.~\ref{fig:lo} also
contribute at the NLO as ${\cal{M}}_{NLO}^{6,7}$ via self-energy diagrams
labelled 6--7 in Fig.~\ref{fig:nlo_self}; see Appendix~\ref{sec:virtual}
for more details.

We can write the NLO virtual contribution as the coefficient of the
${\cal{O}}(\alpha)$ term in 
\begin{eqnarray}
\sigma_{NLO}^{Virtual} & \propto & 
\int \left\vert {\cal{M}}_{LO}^t - {\cal{M}}_{LO}^u +
{\cal{M}}_{NLO}^{1\hbox{--}7,t} -
{\cal{M}}_{NLO}^{1\hbox{--}7,u} \right \vert_\alpha^2~,
\label{eq:nlo_virtual}
\end{eqnarray}
where we have suppressed the phase space and other factors in the interest
of clarity, and the subscript $\alpha$ indicates that we are considering
only the NLO contribution. Each of the higher order $t$-channel matrix elements in
Eq.~\ref{eq:nlo_virtual} gets contributions from the $t$-channel diagrams
shown in Figs.~\ref{fig:nlo} and \ref{fig:nlo_self} (and similarly,
their $u$-channel counterparts).

\begin{figure}[thp]
\begin{center}
\includegraphics[width=0.25\textwidth]{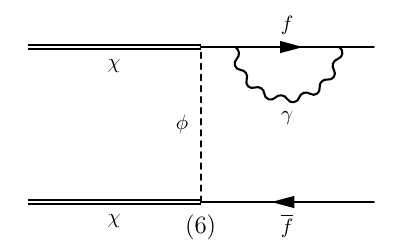}
\includegraphics[width=0.25\textwidth]{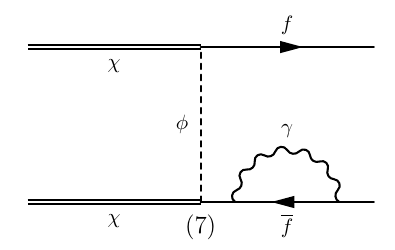}
\caption{\small \em The $t$-channel fermion (L) and anti-fermion (R)
virtual photon self-energy corrections to the dark matter annihilation
process at next to leading order (NLO). Analogous contributions from the
$u$-channel diagrams exist. See Appendix~\ref{sec:virtual} for details.}
\label{fig:nlo_self}
\end{center}
\end{figure}

Since the dark matter particle is assumed to be Majorana, both
the $t$-channel diagrams shown here and their crossed $u$-channel
counterparts, along with the $tu$ cross contributions also exist;
in contrast, only the $t$-channel diagrams contribute when the dark
matter particle is a Dirac fermion. While we have calculated both
the ${\cal{O}}(T^2)$ and ${\cal{O}}(T^4)$ NLO corrections both in the
relativistic case and in the non-relativistic limit using the Grammer
and Yennie approach, we present here the dominant ${\cal{O}}(T^2)$
results. There are three different thermal contributions possible: one
where the the photon propagator contributes through its thermal part,
and the others where the fermion or anti-fermion propagator contribute
through their thermal parts. In what follows, we shall refer to these
as the virtual {\em thermal photon}, {\em thermal fermion} and {\em thermal
anti-fermion} contributions respectively. Note that the case where
both the photon and fermion/anti-fermion are thermal is kinematically
disallowed (see Ref.~\cite{Butola:2024oia} for details).

The thermal virtual contribution at NLO is given by the sum of the thermal
photon, thermal fermion, and thermal anti-fermion contributions. Since
the thermal contribution is already at order ${\cal{O}}(T^2)$, we compute
only the $s$-wave terms since the $p$-wave ones contribute one more power
of temperature to the collision term; see Eq.~\ref{eq:sigma_coll}; hence,
the $s$-wave terms have the most significant contribution at freeze-out
when $m_\chi/T \sim 20$.

It is clear that Diagrams 2, 6, and 7 from Figs.~\ref{fig:nlo}
and \ref{fig:nlo_self} contribute at order ${\cal{O}}(1/m_\phi^4)$
in the scalar mass while Diagrams 1, 3, and 5 contribute at order
${\cal{O}}(1/m_\phi^6)$ while Diagram 4 is highly suppressed, at
order ${\cal{O}}(1/m_\phi^8)$. In the centre-of-momentum frame, with
energy-momentum of the dark matter particles given by $q'^\mu, q^\mu =
(H, 0, 0, \pm Q)$ and that of the fermions given by $p'^\mu, p^\mu =
(H, \pm P \sin\theta, 0, \pm P\cos\theta)$, so that the Mandelstam
variable $s = 4H^2$, we obtain\footnote{The ${\cal{O}}(1/m_\phi^4)$
term, arising from the finite self-energy corrections of Diagrams 6 and
7 shown in Fig.~\ref{fig:nlo_self} and their $u$-channel and crossed
$tu$-channel counterparts, were missed in Ref.~\cite{Butola:2024oia}.}:
\begin{align}
\sigma_{NLO}^{Virtual} (s\hbox{-wave}) & =  
\frac{\alpha \lambda^4}{128} ~\frac{m_f^2 \, T^2}{\sqrt{s} Q}
 \left[-\frac{8}{m_\phi^4} + \frac{32 m_\chi^2}{3 m_\phi^6}
 + \frac{128 m_\chi^4}{3 m_\phi^8} \right] + {\cal{O}}(v^2)~, \nonumber \\
 & = \frac{\alpha \lambda^4}{128} ~\frac{m_f^2 \, T^2}{m_\chi^2 v_{rel}}
 \left[-\frac{8}{m_\phi^4} + \frac{32 m_\chi^2}{3 m_\phi^6}
 + \frac{128 m_\chi^4}{3 m_\phi^8} \right] + {\cal{O}}(v^2)~.
\label{eq:sigma_NLOV}
\end{align}
The NLO virtual thermal cross section was computed in our earlier
work \cite{Butola:2024oia}; the leading NLO contribution was also
calculated earlier by Ref.~\cite{Beneke:2014gla} who explicitly showed
the cancellation of soft and collinear divergences at NLO. The NLO
thermal cross section was found to vanish in Ref.~\cite{Beneke:2014gla}
in the massless limit; this was subsequently explained by them using the
Operator Product Expansion approach in Ref.~\cite{Beneke:2016ghp} where
the cross section for annihilation of Dirac-type dark matter particles
was also computed. In our earlier work, we had used the GY technique
to directly calculate the IR finite $G$ photon contribution and we had
dropped the logarithmic terms that give rise to collinear divergences
since they had been shown \cite{Beneke:2014gla} to cancel. In this paper,
since we have now computed the real photon thermal contributions, we will
explicitly address the issue of cancellation of collinear divergences and
show that the structure of collinear terms is interesting and non-trivial.

Notice the factor $m_f^2$ in all terms due to helicity suppression of
the Majorana dark matter annihilation channel; in contrast, we see below
that the Dirac dark matter annihilation cross section is not helicity
suppressed:
\begin{align}
\sigma_{NLO}^{Dirac, Virtual} (s\hbox{-wave}) & = 
\frac{\alpha \lambda^4}{128} ~\frac{T^2}{m_\chi^2}
 \left[-\frac{4(2 m_\chi^2 + m_f^2)}{m_\phi^4} + \frac{16 m_\chi^2 (2
 m_\chi^2 - m_f^2)}{3 m_\phi^6} \right. \nonumber \\
 & \qquad \qquad \left. + \frac{64 m_\chi^4 ( 2 m_\chi^2 - 3 m_f^2)}{3 m_\phi^8} \right] + {\cal{O}}(v^2)~.
\label{eq:sigma_NLOVDirac}
\end{align}
Before we go on to compute the NLO thermal corrections with real photon
emission/absorption, we outline the Grammer and Yennie technique for
use in the later sections. The informed reader may prefer to skip
to Section~\ref{sec:real}.

\section{Infra-red divergences and the Grammer and Yennie technique}
\label{sec:GY}

We have used the Grammer and Yennie (GY) technique to calculate the IR
finite part of the cross section for both the virtual and real photon
cases. We now summarise this approach which helps simplify the separation
of the soft infra-red (IR) divergences\footnote{We shall see that the
technique fails to factorise out the collinear divergences. However, the
technique is still useful because cancellations occur at the integrand
level and hence there is no need to compute divergent integrals.}. The
GY technique \cite{Grammer:1973db, Yennie:1961ad} was used to factorise
the infra-red (IR) divergences in a zero temperature quantum field
theory. There are additional {\em linear} IR divergences due to photons
in a thermal field theory. Hence the demonstration of cancellation of IR
divergences includes demonstrating the cancellation of both the linear
and (sub-leading) logarithmic divergences in the thermal theory. Note
that there are no additional ultra-violet (UV) divergences because the
number operator in both the thermal propagators and phase space acts as
a damping factor. The GY technique was extended to the case of thermal
field theory, first for fermionic thermal QED \cite{Indumathi:1996ec},
and later for charged thermal scalars and fermions interacting with
dark matter particles with thermal QED corrections \cite{Sen:2020oix,
Sen:2018ybx}, where it was shown that the infra-red divergences cancel
between virtual and real photon insertions, order by order, to all
orders. The technique uses the fact that only the photon contributions
are IR divergent, while fermion insertions do not lead to additional IR
divergences. The IR divergences cancel between virtual and real photon
insertions as discussed below.

Consider first the insertion of a {\em virtual photon} with
momentum $k$ into a lower order graph. The procedure starts with writing
the virtual photon propagator as the sum of two parts, the so-called $K$
and $G$ photons:
\begin{align} \nonumber -i
g^{\mu{\nu}} & \to {-i}
	\left\{\strut \left[\strut g^{\mu\nu} - b_k (p_f, p_i)\,k^\mu
	k^\nu \right] + \left[\strut b_k (p_f, p_i) k^\mu k^\nu \right]
	\right\}~, \nonumber \\
 & \equiv {-i} \left\{\left[\strut G_k^{\mu\nu}\right] +
 \left[K_k^{\mu\nu}\right] \right\}~.
\label{eq:KG}
\end{align}
The IR divergence is completely contained in the $K$ photon contribution
while the $G$ photon contribution is IR finite. The notation $p_f$
and $p_i$ is used to separate the final and initial ``legs'' (where the
photon is inserted at vertices $\mu, \nu$) which are defined as being
to the right or left of the special vertex $V$ (see Fig.~\ref{fig:lo})
at which the momentum $q'$ is guaranteed to be hard (not soft). Hence,
for the $t$-channel diagram in Fig.~\ref{fig:lo}, the fermion line with
$p_f=p'$ is defined as the final ``leg'' while the anti-fermion and scalar
lines together form the initial leg with $p_i = p$; for the $u$-channel
diagrams, the scalar forms a part of the $p'$ leg instead. Then the
factor $b_k$ is given by
\begin{align}
b_k(p_f,p_i) = \frac{1}{2} \left[ \frac{(2p_f-k) \cdot
(2p_i-k)}{((p_f-k)^2-m^2)((p_i-k)^2-m^2)} + (k \leftrightarrow -k)
\right]~,
\label{eq:bk}
\end{align}
is defined symmetrically in $k\to -k$ for the thermal case (in contrast to
the original definition in Ref.~\cite{Grammer:1973db}), and is a function
of $k$ as well as the momenta, $p_f$, $p_i$. It can then be shown that
the $K$ photon insertions contain the IR divergent pieces while the $G$
photon contribution is IR finite. The proof is long; but the key point
is that the insertion of an additional virtual $K$ thermal photon with
momentum $k$ in all possible ways into a lower ($n^{\rm th}$) order graph
results in the following factorisation of the consequent $(n+1)^{\rm
th}$ order matrix element in terms of the lower $n^{\rm th}$ order 
matrix element:

\begin{align} \nonumber
{\cal{M}}_{n+1}^{K{\rm photon, tot}} 
 & = \left[\frac{ie^2}{2} \int
	\frac{{\rm d}^4 k}{(2\pi)^4} \, D^{11} (k) \,
	\left[\strut b_k(p',p') - 2 b_k(p',p) + b_k(p,p) \right]
	\rule{0pt}{16pt} \right]
	 {\cal{M}}_{n}~, \nonumber \\
	 & \equiv \left[B \right] {\cal{M}}_{n}~,
\label{eq:K}
\end{align}
where $D^{11}(k)$ is the $(11)$ element of the photon propagator
(excluding the tensor $g^{\mu\nu}$ term); see Eqs.~\ref{eq:sprop},
\ref{eq:thermalD}, and its $T$-dependent part contains the number
operator, $n_B(|k^0|)$. Notice that the prefactor $B$ contains the entire
$k$ dependence and is IR divergent.


While the factorisation of the IR divergent part occurs in the {\em
matrix element} for virtual photon insertions, it occurs in the {\em
square} of the matrix elements for the real photon case. Note that
both emission and absorption of real photons into/from the heat bath
are possible in the thermal case; in fact, this is essential in order
to show the IR divergence cancellation between virtual and real photon
contributions. When a real photon with momentum $k$ is inserted into
a lower order diagram at a vertex $\mu$, the polarisation sum in the
squared matrix element can be re-written in terms of the $\widetilde{K}$
and $\widetilde{G}$ contributions:
\begin{equation}
\begin{array}{rcl}
\displaystyle \sum\limits_{\rm pol} \epsilon^{\mu *} (k)\,\epsilon^\nu (k)
= \displaystyle - g^{\mu\nu}
 & = & \displaystyle -\left\{\strut \left[\strut g^{\mu\nu} -
	\tilde{b}_k(p_f,p_i) k^\mu k^\nu \right] +
	\left[\strut \tilde{b}_k(p_f,p_i) k^\mu k^\nu \right] \right\}~,
	 \\[2ex]
 & \equiv & \displaystyle -\left\{\left[ \widetilde{G}_k^{\mu\nu}\right] + \left[
 \widetilde{K}_k^{\mu\nu}\right]\right\}~,
\end{array}
\label{eq:KGtilde}
\end{equation}
with $\tilde{b}_k(p_f, p_i) = b_k(p_f, p_i)\vert_{k^2 = 0}$ for real
photons. Again, $p_i$ ($p_f$) is the momentum $p'$ or $p$ depending on
whether the real photon insertion was on the $p'$ or $p$ leg in the
$n^{\rm th}$ order matrix element ${\cal M}_{n+1}$ (or its conjugate
${\cal M}_{n+1}^\dagger$). As in the case for virtual $K$ photon
insertion, the insertion of a $\widetilde{K}$ real photon into an $n^{\rm
th}$ order graph leads to a cross section that is proportional to the
lower order one by an overall factor analogous to that in Eq.~\ref{eq:K}
for virtual photon insertion:
\begin{align} \nonumber
\left \vert {\cal{M}}_{n+1}^{\widetilde{K}\gamma,{\rm tot}} \right \vert^2
	& \propto -e^2 \left[\strut \tilde{b}_k(p,p) -2
	\tilde{b}_k(p',p) + \tilde{b}_k(p',p') \right]
\left \vert {\cal{M}}_{n}^{{\rm tot}} \right \vert^2~.
\label{eq:ktilde}
\end{align}
Since we are dealing with real photon insertions, recall that the phase
space factor also contains $k$ dependence:
\begin{equation}
\int {\rm d} P_k \equiv \int \frac{{\rm d}^4 k}{(2\pi)^4} 2\pi \delta(k^2)
 \left[\theta(k^0) (1 + n_B(\vert k^0 \vert)) + \theta(-k^0)
 n_B(\vert k^0 \vert) \right]~,
\label{eq:realphspace}
\end{equation}
where the two terms account for both the probability of emission into
a heat bath at temperature $T$ (proportional to $(n_B(\vert k^0 \vert)
+1)$) and absorption from the heat bath (proportional to $n_B(\vert
k^0 \vert)$). Then the total $k$-dependent part that factorises out of
the expression can be written as \cite{Sen:2020oix, Sen:2018ybx},
\begin{equation}
\widetilde{B}(x) = -e^2 \int {\rm d}P_k \left[\strut \tilde{b}_k(p,p) -2
	\tilde{b}_k(p',p) + \tilde{b}_k(p',p') \right]
	\exp\left[ {\pm i\, k \cdot x}\right]~,
\label{eq:Btilde}
\end{equation}
where the $\pm$ sign depends on whether the photon with momentum $k$
is emitted/absorbed. After some simplification, and including the finite
contributions from the $G$ and $\widetilde{G}$ insertions, the total
cross section can be expressed as
\begin{align}
\sigma^{\rm tot} & = 
	\int {\rm d} ^4x \, e^{-i(q'+q-p'-p)\cdot x}\, {\rm d}P_{p'}
	{\rm d}P_{p} \exp\left[ B+B^*+\widetilde{B} \right] \,
	\sigma^{\rm finite} (x)~.
\label{eq:sigmatot}
\end{align}
Here $\sigma^{\rm finite}$ contains the finite $G$ and $\widetilde{G}$
photon contributions from the virtual and real thermal photons
respectively, as well as the (finite) thermal fermion contributions.
In the limit $k \to 0$, the exponential IR divergent parts of
both the virtual and real photon contributions can be seen to cancel:
\begin{align} 
(B+B^*) + \widetilde{B} 
 & \stackrel{k \to 0}{\longrightarrow} ~0 + {\cal O}(k^2)~.
\label{eq:finite}
\end{align} 
Hence the total cross section is IR finite to all orders, with the IR
divergent part of the real photon cross section cancelling against
the corresponding IR divergent part of the virtual contribution. We
will use this approach to compute the real thermal corrections to dark
matter annihilation processes. In the next section, we will therefore
compute the finite remainder at ${\cal{O}}(\alpha)$ arising from the
finite combination, $[(B+B^*) + \widetilde{B}]$, which we shall label,
for convenience, as $(K + \widetilde{K})$, and the finite contribution,
$\widetilde{G}$. With this summary explanation of the GY technique as
well as the discussion on the virtual photon NLO thermal corrections in
the previous section, we now go on to the main results of this paper.

\section{The real photon thermal contribution to the dark matter
annihilation cross section}
\label{sec:real}

\begin{figure}[htp]
\centering
\includegraphics[width=0.32\textwidth]{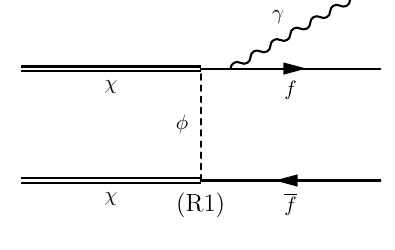}
\includegraphics[width=0.32\textwidth]{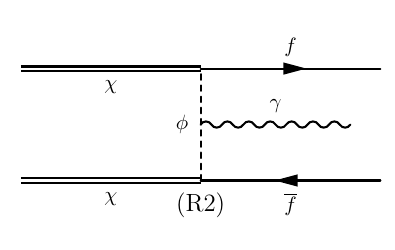}
\includegraphics[width=0.32\textwidth]{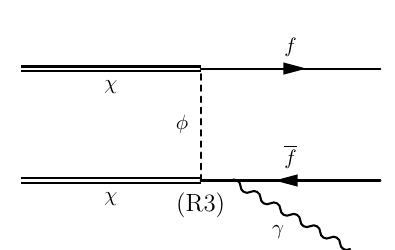}
\caption{\small \em The $t$-channel real photon emission diagrams
contributing to the dark matter annihilation process at next to leading
order (NLO). Diagrams are labelled from $R1$--$R3$. Similar terms contribute
when the photon is instead absorbed from the heat bath. Analogous
contributions from the $u$-channel diagrams also exist.}
\label{fig:nlo_real}
\end{figure}

The thermal contributions to $\chi \chi \to f \overline{f}
(\gamma)$ arise from insertions of real photons into the set of LO
diagrams shown in Fig.~\ref{fig:lo}, which can be both emitted into or
absorbed from the heat bath at temperature $T$. The relevant $t$-channel
diagrams $R1, R2, R3$ are shown in Fig.~\ref{fig:nlo_real} while
the corresponding $u$-channel diagrams $R4, R5, R6$ are obtained by
crossing these diagrams.

In contrast to the virtual photon case where the thermal
${\cal{O}}(\alpha)$ contribution arose from the product of the NLO
(virtual) matrix elements of the diagrams in Fig.~\ref{fig:nlo} with the
LO matrix elements in Fig.~\ref{fig:lo}, the ${\cal{O}}(\alpha)$ thermal
contribution in the real case arises from the {\em square} of the matrix
elements corresponding to the diagrams in Fig.~\ref{fig:nlo_real}.

A quick look at the diagrams in Fig.~\ref{fig:nlo_real} is sufficient
to realise that all particles contribute through their type-1
thermal fields at every vertex. As defined in Appendix~\ref{sec:tft},
propagators (for photons, scalars and fermions) are given as the sum of a
temperature-independent part and a temperature dependent part, the latter
of which is on-shell and weighted by the appropriate (Fermi Dirac or
Bose Einstein) distribution function. It can be seen that neither of the
fermion or anti-fermion propagators can contribute through their thermal
parts since three particles at a vertex are kinematically forbidden to
be all on-shell. Hence, all thermal contributions to the dark matter
annihilation process with real photon emission/absorption arise from
the corresponding phase space elements. For a particle with momentum $p$
and mass $m_p$, the phase space factor is given by
\begin{align}
dP_p = \int \frac{{\rm d}^4 p}{(2\pi)^4} \left(
2\pi \delta(p^2 -m_p^2) \right) \left[ \theta(p^0) \pm n(\vert p^0
\vert) \right]~.
\label{eq:phase}
\end{align}
Here the second term is the thermal contribution, proportional to $n
\equiv n_B \,(n_F)$, the Bose (Fermi) distribution function, for bosons
and fermions respectively, with the $+\,(-)$ sign corresponding to bosons
(fermions). In the calculation that follows, only one of the particles
(photon, fermion, anti-fermion) at a time is taken to contribute via its
thermal part, since the cross section is otherwise suppressed by products
of distribution functions. We therefore have three contributions from each
diagram in Fig.~\ref{fig:nlo_real} (and its $u$-channel counterparts):
when each of the real photon, fermion, or anti-fermion contributes via the
thermal part of its phase space. We shall refer to them as the {\em real 
thermal photon}, {\em thermal fermion}, and {\em thermal anti-fermion}
contributions respectively.

We apply the Grammer and Yennie technique \cite{Grammer:1973db} as
explained in Section~\ref{sec:GY}. Hence each of the thermal photon
contributions can be separated into $\widetilde{K}$ and $\widetilde{G}$
parts by applying the GY technique. However, it turns out that
while the GY technique isolates the {\em soft} IR divergence into the
$\widetilde{K}$ (and $K$) photon contribution, it fails to separate the
{\em collinear} divergence the same way. Hence, in what follows, we shall
first demonstrate the cancellation of the soft IR divergences between
the $\widetilde{K}$ and $K$ photon contributions, while suppressing
the collinear terms, and deal with the collinear divergences
separately, in a different section, Section~\ref{ssec:collinear},
where we show explicitly that these logarithmic terms cancel between
real and virtual contributions. It therefore appears that the use of
the GY technique is restricted to those processes where the collinear
divergences are independently known to factorise and cancel between
virtual and real contributions, order by order in the theory. Even so,
there is an advantage in using the GY technique since the divergences
cancel at the {\em integrand} level and simplify the calculation.

\subsection{Kinematics of \, $\chi \chi \to f \, \overline{f} \, (\gamma)$}

We consider the real photon emission/absorption process in the centre of
momentum frame where the dark matter particles have common energy-momentum
$q', q = (H, 0, 0, \pm Q)$. We use the same kinematics as in the virtual
case, except for the fact that physical momentum $k$ is gained or lost
in the real photon process, so that the fermion (with momentum $p'$) and
anti-fermion (with momentum $p$) no longer have the same energy or
momentum (magnitude). The real photon cross section is given by the sum
of 21 contributions as follows:
\begin{align}
\sigma_{NLO}^{Real} & = 
\left[
\sum_{i=1}^6 \sigma_{i,i}^{Real} + (-1)^{S_p} \sum_{i<j}^6 \sigma_{i,j}
^{Real}\right]~,
\label{eq:sigma_nlo_real}
\end{align}
where the individual terms are defined in terms of the squares of
various matrix elements:
\begin{align}
\sigma_{i,i}^{Real} & = \frac{1}{4 \sqrt{s} Q} \int {\rm d}P_{p'}
{\rm d}P_{p} {\rm d}P_{k} (2\pi)^4 \delta^4
(q'\!+\!q\!-\!p'\!-\!p\!-\!k)\!
\left\vert M_{R_i} \right\vert^2~, \nonumber \\
\sigma_{i,j}^{Real} & = \frac{1}{4 \sqrt{s} Q} \int {\rm d}P_{p'}
{\rm d}P_{p} {\rm d}P_{k} (2\pi)^4 \delta^4
(q'\!+\!q\!-\!p'\!-\!p\!-\!k)\! \left\vert M_{R_i} M_{R_j}^\dagger + h.c.
\right\vert^2~,
\label{eq:sigma_m2}
\end{align}
with $ -\infty \le k^0 \le \infty$, thus allowing for both emission and
absorption of the real photon, and
\begin{align}
S_p & = 0~; ~~ i,j \in \{1,2,3\} \hbox{ or } \{4,5,6\}~, \nonumber \\
	& = 1~; ~~ i \in \{1,2,3\} \hbox{ and } j \in \{4,5,6\}~.
\label{eq:nlo_Sp}
\end{align}
Here the phase space factors $dP_i$ are defined in Eq.~\ref{eq:phase},
and the matrix elements correspond to the three $t$-channel diagrams $R1,
R2, R3$ as seen in Fig.~\ref{fig:nlo_real} and $R4, R5, R6$ are their
$u$-channel counterparts. Therefore $\sigma_{i,j}$, $i,j \in 1$--3,
are from the $t$-channel diagrams, $\sigma_{i,j}$, $i,j \in 4$--6,
are from the $u$-channel diagrams, and the remaining from the crossed
$tu$-channel diagrams. In addition, each of these have contributions
from three distinct sets of terms: when the photon is thermal, when
the fermion is thermal, and when the anti-fermion is thermal. That is,
we can write
\begin{align}
\sigma_{i,j}^{Real} & = 
\sigma_{i,j}^{\gamma, Real} + \sigma_{i,j}^{f, Real} +
\sigma_{i,j}^{\overline{f}, Real}~,
\label{eq:sigma_gffbar}
\end{align}
for all values of $(i,j)$, where the contributions correspond to
including the thermal part of the phase space in Eq.~\ref{eq:sigma_m2}
for the photon, fermion and anti-fermion respectively.

The frame used for the computation is as follows: of the two
non-thermal particles, one is integrated out using the energy-momentum
conserving delta function shown in Eq.~\ref{eq:sigma_m2}. The other
one is rotated into the $z$-direction with energy-momentum $(E',0,0,
P')$, while the thermal particle is arbitrarily aligned in the direction
$\hat{n}'(\Omega')$ defined by the angular coordinates $(\theta',
\phi')$. This allows for the simplest kinematics, and yields, for
instance, when the {\em photon is thermal} with momentum $k^\mu =
(k^0, K \hat{n}')$, the anti-fermion momentum is integrated out, and the
fermion is rotated into the $z$-axis:
\begin{align}
\sigma_{i,j}^{\gamma, Real}
& = \frac{1}{32 \sqrt{s} Q} \frac{1}{2(2\pi)^3} \int_0^{\omega_{max}}
\hspace{-0.5cm} {\rm d} \omega \int_{E'_{min}}^{E'_{max}} {\rm d} E'
\int_{-1}^1 \!\!{\rm d}\!\cos\theta \int_{0}^{2\pi} {\rm d} \phi'
\nonumber \\
 & \qquad \qquad \times ~n_B(\omega) \left[ F_{i,j}^+(\omega, \omega,
 \Omega') + F_{i,j}^-(-\omega, \omega, \Omega') \right]~,
\label{eq:nlo_Tgamma}
\end{align}
where $\omega = \vert k^0 \vert$, the factor $n_B(\omega)$ has been
taken out of the integrand to show the dependence on the distribution
function, and only the $k^\mu$ dependence of the integrand,
$F^\pm_{i,j}(k^0, K, \Omega')$, has been indicated. Here $\theta$ is
the angle between the (rotated) $z$-axis and the dark matter momentum
direction; while the thermal photon angle $\theta'$ is given by
\begin{equation}
\cos\theta' = \frac{(s - 2 \sqrt{s}
( E' \pm \omega) \pm 2 E' \omega)}{2 P' \omega}~.
\label{eq:Tgammatheta}
\end{equation}
The $\pm$ signs correspond to the case of emission/absorption of the
photon. The limits on $E'$ are obtained from constraints on $\vert
\cos\theta' \vert < 1$ while $\omega_{max} = (s-m_f^2)/(2\sqrt{s})$.

When the {\em thermal fermion} contribution is under consideration, the
anti-fermion momentum is integrated out as in the thermal photon case,
and the photon is rotated into the $z$-axis, so that $k^\mu = (E', 0,
0, E')$. In this case, the fermion has momentum $p'^\mu = (p'^0, K_t
\hat{n}')$, and we get an expression analogous to Eq.~\ref{eq:nlo_Tgamma}
for $\sigma_{i,j}^{f, Real}$, while also accounting for the fact that
the distribution function $n_B(\omega)$ in Eq.~\ref{eq:nlo_Tgamma} is to
be replaced by $(-n_F(\omega_t))$ as per the phase space definition in
Eq.~\ref{eq:phase}; furthermore, the fermion has non-zero mass so that
the thermal fermion angle $\theta'$ is given by
\begin{equation}
\cos\theta' = \frac{(s - 2 \sqrt{s}
( E' \pm \omega_t) \pm 2 E' \omega_t)}{2 E' K_t}~,
\label{eq:Tftheta}
\end{equation}
where the thermal fermion energy and momentum magnitude are represented
by $\omega_t = \vert p'^0 \vert$, and $K_t$ with $\omega_t^2 = K_t^2 +
m_f^2$ to distinguish them from the massless photon contribution. 
The case for thermal anti-fermion merely exchanges the fermion and
anti-fermion momenta and is straightforward.

In all cases, it can be seen from Eq.~\ref{eq:nlo_Tgamma} that high
energy contributions are suppressed due to the presence of the appropriate
distribution functions,
\begin{align}
n_B(\omega) & = \frac{1}{\exp[\beta \omega] - 1}~, \nonumber \\
n_F(\omega_t) & = \frac{1}{\exp[\beta \omega_t] + 1}~.
\label{eq:nBF}
\end{align}
Hence, the upper limit on $\omega$ ($\omega_t$) can be taken to be
infinity so that these integrals can be analytically performed. The
leading thermal contribution of order ${\cal{O}}(T^2)$ then arises from
the integrals:
\begin{align}
\int_0^\infty \omega \, {\rm d} \omega \, n_B(\omega) & = \frac{\pi^2 T^2}{6}~,
\nonumber \\
\int_0^\infty \omega_t \, {\rm d} \omega_t \, n_F(\omega_t) & = \frac{\pi^2
T^2}{12}~,
\end{align}
where we have assumed the massless limit for the thermal fermion
integration.

\subsection{The matrix elements}

The relevant matrix elements for the real photon emission/absorption
process $\chi \chi \to f \overline{f} (\gamma)$ corresponding to the
diagrams $R1$--$R3$ shown in Fig.~\ref{fig:nlo_real} are 
\begin{align}
M_{R1} & = \frac{i e \vert\lambda \vert^2}{2 k \cdot p' ~l_\Phi}
\left[\overline{v}(q,m_\chi) P_L \,v(p,m_f)\right]
\left[\overline{u} (p',m_f) \gamma_\mu \left(\slashed{k}+\slashed{p}'+
m_f\right) P_R \, u(q',m_\chi) \right] \epsilon_k^{*\mu}~; \nonumber \\
M_{R2} & = \frac{i e \vert\lambda \vert^2}{lk_\Phi ~l_\Phi}
\left(2 l - k \right)_\mu \left[\overline{v}(q,m_\chi) P_L
 v(p,m_f) \right] \left[ \overline{u}(p',m_f) P_R \, u(q',m_\chi)
 \right] \epsilon_k^{*\mu}~; \nonumber \\
M_{R3} & = \frac{i e \vert\lambda \vert^2}{2 k \cdot p ~lk_\Phi}
\left[\overline{v}(q,m_\chi) P_L \left(-\slashed{k}-\slashed{p}+m_f\right)
\gamma_\mu \,v(p,m_f)\right]
\left[\overline{u} (p',m_f) P_R \, u(q',m_\chi) \right]
\epsilon_k^{*\mu}~.
\label{eq:matrix}
\end{align}
Here $P_{R,L} = (1 \pm \gamma_5)/2$, $l_\Phi \equiv (l^2 -m_\phi^2)$
arises from the propagator of the scalar with momentum $l =
q -p$. Similarly, $lk_\Phi = (l-k)^2 - m_\phi^2$. The $u$-channel
matrix elements can be obtained by crossing, with momentum $l' = q' -p$
replacing $l$.

It can be seen from Eq.~\ref{eq:matrix} that the squared matrix elements
contain products of the photon polarisation tensor, $\epsilon_k^\mu$,
to which the GY-defined polarisation sum shown in Eq.~\ref{eq:KGtilde}
can be applied in order to separate the $\widetilde{K}$ and
$\widetilde{G}$ contributions in the thermal photon case. Since there
are known to be no IR divergences \cite{Sen:2020oix, Sen:2018ybx}
in the thermal fermion contributions, the standard polarisation sum,
\begin{align}
\sum\limits_{\rm pol} \epsilon_k^{\mu *}\,\epsilon_k^\nu
= - g^{\mu\nu}~,
\end{align}
is used in this case.

The angular integrals were performed using FeynCalc 10.0.0
\cite{Shtabovenko:2023idz, Mertig:1990an} software
with Mathematica 13.1 \cite{Mathematica}. We will first discuss the soft
IR and collinear divergences in the next sections, before we consider
the finite contribution. The soft IR divergences arise from the
$\widetilde{K}$ photon contributions. Since only thermal photons
contribute to these terms, we now consider the case where the thermal
part of the photon phase space (see Eq.~\ref{eq:phase}) contributes.

\subsection{The thermal photon contribution and IR divergences}

The soft IR divergences are known to cancel between the real and virtual
thermal photon contributions. These are contained in the {\em thermal
photon} contributions while the {\em thermal fermion} contributions are
IR finite due to the nature of the Fermi distribution function; see
Eq.~\ref{eq:nBF}. As shown in Refs.~\cite{Sen:2020oix, Sen:2018ybx},
the soft IR divergences can be isolated into the virtual $K$ and real
$\widetilde{K}$ photon contributions as defined in Eqs.~\ref{eq:KG},
\ref{eq:KGtilde}. The sum of these two contributions is finite (this was
first shown to NLO in Ref.~\cite{Beneke:2014gla}); we will explicitly
show below the cancellation of the soft IR divergences between these two
contributions, and also compute the finite remainder. Some details of
the virtual NLO thermal correction to $\chi \chi \to f \overline{f}$ are
given for convenience in Appendix~\ref{sec:virtual}; for more details,
see our earlier work, Ref.~\cite{Butola:2024oia}.

\subsubsection{The $\widetilde{K}$ contribution and soft IR divergences}

The soft divergence lies in the $\omega$ integrals of the {\em thermal
photon} contribution to the real photon cross section which we label
$\sigma_{i,j}^{\gamma, Real}$ (see Eq.~\ref{eq:nlo_Tgamma}); hence
we integrate out the angular variables, leaving only the $\omega$
dependence. The $\widetilde{K}$ part of the thermal real photon cross
section can be obtained by taking the $\widetilde{K}$ part of the
polarisation sum (see Eq.~\ref{eq:KGtilde}) in the expressions for
$\sigma_{i,j}^{\gamma,Real}$. We have
\begin{align}
\sigma_{i,j}^{\gamma, Real}
& = \frac{e^2 \pi \vert \lambda\vert^4}{32 \sqrt{s} Q} \frac{1}{2(2\pi)^3}
\int {\rm d} \omega ~ n_B(\omega) ~ Int^{\gamma}_{i,j}~, \nonumber \\
Int^\gamma_{i,j} & = Int^{\widetilde{K}}_{i,j} + Int^{\widetilde{G}}_{i,j}~,
\label{eq:IntKG}
\end{align}
where $Int^{\widetilde{K}}_{i,j}$ is listed for all $(i,j)$
contributions in Table~\ref{tab:Ktilde}. Here we have retained only
terms upto order ${\cal{O}}(\omega, m_f^2)$ since the term linear
in $\omega$ will give rise to order ${\cal{O}}(T^2)$ corrections
to the cross section. Hence the terms proportional to $1/\omega$ in
Table~\ref{tab:Ktilde} are linearly divergent in the IR. These are
expected to cancel against the corresponding contributions from the
virtual NLO terms.

\begin{table}[htp]
\centering
\begin{tabular}{|c|c|c|c|}
\hline
$i,j$ \rule{0pt}{16pt}
& {\bf Real} $Int^{\widetilde{K}}_{i,j}$ from $Int^\gamma_{i,j}$ 
& {\bf Virtual} $Int_{i,j}^{K_{div}}$ 
& $Int^{\widetilde{K}_{fin}}_{i,j} = 
Int^{\widetilde{K}}_{i,j} + Int_{i,j}^{K_{div}}$ 
\\ \hline
$1,1$ \rule{0pt}{16pt}
& $-m_\chi^2(\frac{32 m_\chi^2 -
8 m_f^2}{\omega m_\phi^4}) -
\omega (\frac{32 m_\chi^2 - 8 m_f^2}{m_\phi^4}) $ 
 &
$m_\chi^2(\frac{32 m_\chi^2 -
8 m_f^2}{\omega m_\phi^4} )$ 
 & 
$-\omega ( \frac{32 m_\chi^2 - 8 m_f^2}{m_\phi^4}
)$
\\ \hline
$2,2$ \rule{0pt}{16pt}
& $-\omega m_\chi^4(\frac{128 m_\chi^2 +
96 m_f^2}{m_\phi^8} )$ & -- & 
$-\omega m_\chi^4(\frac{128 m_\chi^2 +
96 m_f^2}{m_\phi^8} )$ 
\\ \hline
$3,3$ \rule{0pt}{16pt}
& $-m_\chi^2(\frac{32 m_\chi^2 -
8 m_f^2}{\omega m_\phi^4} ) -
\omega ( \frac{32 m_\chi^2 - 8 m_f^2}{m_\phi^4}
)$ 
 &
$m_\chi^2(\frac{32 m_\chi^2 -
8 m_f^2}{\omega m_\phi^4} )$ 
 & 
$-\omega (\frac{32 m_\chi^2 - 8 m_f^2}{m_\phi^4})$
\\ \hline
$4,4$ \rule{0pt}{16pt}
& $-m_\chi^2(\frac{32 m_\chi^2 -
8 m_f^2}{\omega m_\phi^4} ) -
\omega ( \frac{32 m_\chi^2 - 8 m_f^2}{m_\phi^4})$ 
 &
$m_\chi^2(\frac{32 m_\chi^2 -
8 m_f^2}{\omega m_\phi^4} )$ 
 & 
$-\omega (\frac{32 m_\chi^2 - 8 m_f^2}{m_\phi^4})$
\\ \hline
$5,5$ \rule{0pt}{16pt}
& $-\omega m_\chi^4(\frac{128 m_\chi^2 +
96 m_f^2}{m_\phi^8} )$ & -- & 
$-\omega m_\chi^4(\frac{128 m_\chi^2 +
96 m_f^2}{m_\phi^8} )$ 
\\ \hline
$6,6$ \rule{0pt}{16pt}
& $-m_\chi^2(\frac{32 m_\chi^2 -
8 m_f^2}{\omega m_\phi^4} ) -
\omega ( \frac{32 m_\chi^2 - 8 m_f^2}{m_\phi^4}
)$ 
 &
$m_\chi^2(\frac{32 m_\chi^2 -
8 m_f^2}{\omega m_\phi^4} )$ 
 & 
$-\omega (\frac{32 m_\chi^2 - 8 m_f^2}{m_\phi^4})$
\\ \hline
$1,2$ \rule{0pt}{16pt}
 & 
0 \rule{0pt}{16pt}
 & 
0 
 &
 0
\\ \hline
$1,3$ \rule{0pt}{16pt}
 & $\omega ( \frac{32 m_\chi^2 - 32 m_f^2}{m_\phi^4}
)$ 
 &
0 &
$\omega ( \frac{32 m_\chi^2 - 32 m_f^2}{m_\phi^4}
)$ 
\\ \hline
$1,4$ \rule{0pt}{16pt}
 & $-m_\chi^2(\frac{64 m_\chi^2 -
48 m_f^2}{\omega m_\phi^4}) -
\omega ( \frac{64 m_\chi^2}{m_\phi^4}
)$ 
 &
 $m_\chi^2(\frac{64 m_\chi^2 - 48 m_f^2}{\omega
 m_\phi^4} )$
 &
$-\omega ( \frac{64 m_\chi^2}{m_\phi^4} )$ 
\\ \hline
$1,5+ 2,4$ \rule{0pt}{16pt}
 & 
$\omega m_\chi^2 ( \frac{256 m_\chi^2 - 96
m_f^2}{m_\phi^6} )$ 
& 
--
& 
$\omega m_\chi^2 ( \frac{256 m_\chi^2 - 96
m_f^2}{m_\phi^6} )$ 
\\ \hline
$1,6+3,4$ \rule{0pt}{16pt}
 & 0
 &
0 \rule{0pt}{16pt}
 & 0
\\ \hline
$2,3$ \rule{0pt}{16pt}
 & $\omega m_\chi^2(\frac{256 m_\chi^2 -
32 m_f^2}{m_\phi^6} )$ 
 &
 0
 & $\omega m_\chi^2(\frac{256 m_\chi^2 -
32 m_f^2}{m_\phi^6} )$ 
\\ \hline
$2,5$ \rule{0pt}{16pt}
 & $\omega m_\chi^4(\frac{512 m_\chi^2 -
768 m_f^2}{m_\phi^8} )$ 
 &
 --
 & $\omega m_\chi^4(\frac{512 m_\chi^2 -
768 m_f^2}{m_\phi^8} )$ 
\\ \hline
$2,6+3,5$ \rule{0pt}{16pt}
 & $\omega m_\chi^2(\frac{256 m_\chi^2 -
96 m_f^2}{m_\phi^6} )$ 
 &
 0
 & 
 $\omega m_\chi^2(\frac{256 m_\chi^2 -
96 m_f^2}{m_\phi^6} )$ 
\\ \hline
$3,6$ \rule{0pt}{16pt}
 & $-m_\chi^2(\frac{64 m_\chi^2 -
48 m_f^2}{\omega m_\phi^4}) -
\omega ( \frac{64 m_\chi^2}{m_\phi^4}
)$ 
 &
 $m_\chi^2(\frac{64 m_\chi^2 - 48 m_f^2}{\omega
 m_\phi^4} )$
 &
$-\omega ( \frac{64 m_\chi^2}{m_\phi^4} )$ 
\\ \hline
$4,5$ \rule{0pt}{16pt}
 & $\omega m_\chi^2(\frac{256 m_\chi^2 -
32 m_f^2}{m_\phi^6} )$ 
 &
 0
 & 
 $\omega m_\chi^2(\frac{256 m_\chi^2 -
32 m_f^2}{m_\phi^6} )$ 
\\ \hline
$4,6$ \rule{0pt}{16pt}
 & $\omega ( \frac{32 m_\chi^2 - 32 m_f^2}{m_\phi^4}
)$ 
 &
0 &
$\omega ( \frac{32 m_\chi^2 - 32 m_f^2}{m_\phi^4}
)$ 
\\ \hline
$5,6$ \rule{0pt}{16pt}
& 0 \rule{0pt}{16pt}
& 0
& 0 
\\ \hline
\end{tabular}
\caption{\small \em The $\widetilde{K}$ real photon contribution
$Int^{\widetilde{K}}_{i,j}$ (see Eq.~\ref{eq:IntKG}) from various
diagrams when the {\em photon is thermal}. It can be seen that the
IR divergent part of the $\widetilde{K}$ contribution exactly cancels
against the divergent part of the corresponding $K$ virtual thermal photon
contribution, $Int^{K_{div}}_{i,j}$, leaving a finite remainder, which
we label as $Int^{\widetilde{K}_{fin}}_{i,j}$. See text for details.}
\label{tab:Ktilde}
\end{table}

\subsubsection{The real--virtual correspondence}
\label{sssection:corresp}

\begin{figure}[htp]
\centering
\includegraphics[width=0.32\textwidth]{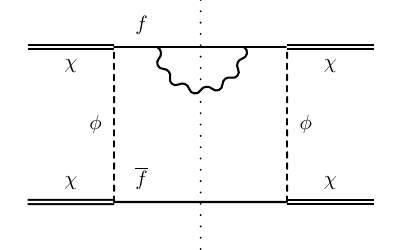}
\includegraphics[width=0.32\textwidth]{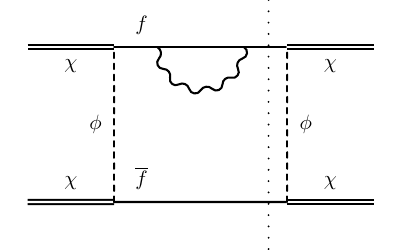}
\caption{\small \em The correspondence between $\sigma_{1,1}^{Real}$ (L)
and $\sigma_{1,1}^{Virtual}$ (R) shown diagrammatically through cut
diagrams.}
\label{fig:cut11}
\end{figure}

In order to determine the virtual terms corresponding to the real
ones, we return to the definition of the real cross section in
Eqs.~\ref{eq:sigma_nlo_real} and \ref{eq:sigma_m2}. The correspondence
is as follows. Consider, for example, the $\sigma_{1,1}^{Real}$
term which arises from the square of the diagram $R1$ shown in
Fig.~\ref{fig:nlo_real} and can be represented by the cut diagram on the
left in Fig.~\ref{fig:cut11}. The corresponding virtual contribution is
then found by shifting the cut on this diagram such that the photon is
virtual as shown on the right side of Fig.~\ref{fig:cut11}. We label
this contribution as $\sigma_{i,j}^{Virtual}$, $(i,j) = (1,1)$, which
is given by (see Eq.~\ref{eq:nlo_virtual})
\begin{align}
\sigma_{1,1}^{Virtual} & \propto 
\int \left[\left({\cal{M}}_{LO}^t\right)^\dagger 
{\cal{M}}_{NLO}^{6,t} + h.c. \right]~,
\label{eq:11correspond}
\end{align}
which arises from the fermion self-energy term shown as Diagram 6
in Fig.~\ref{fig:nlo_self} multiplied by the conjugate of the the LO
$t$-channel diagram in Fig.~\ref{fig:lo}. (The $h.c.$ term is obtained
by moving the cut to the left of the self energy insertion.)

\begin{figure}[htp]
\centering
\includegraphics[width=0.32\textwidth]{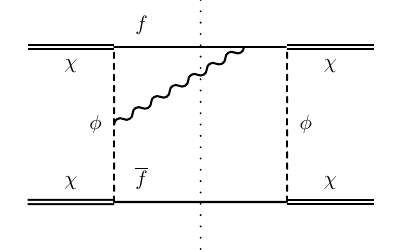}
\includegraphics[width=0.32\textwidth]{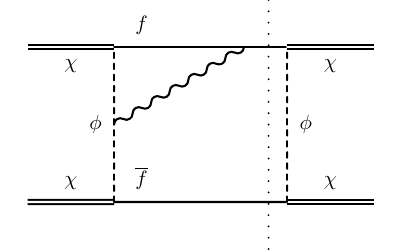}
\caption{\small \em As in Fig.~\ref{fig:cut11} to demonstrate the
$(1,2)$ real--virtual correspondence through cut diagrams.}
\label{fig:cut12}
\end{figure}

Similarly, the virtual counterparts of the $(1,2)$ and $(1,3)$ terms
arise from the contributions of the NLO diagrams labelled (1) and (2)
repectively in Fig.~\ref{fig:nlo} multiplied by the conjugate of the LO
$t$-channel diagram in Fig.~\ref{fig:lo}, etc. The $(1,2)$ correspondence
is shown in Fig.~\ref{fig:cut12}.

We use the results from our earlier work \cite{Butola:2024oia} to obtain
the $K$ photon IR divergent parts of the various virtual thermal photon
diagrams. These are listed as $Int_{i,j}^{K_{div}}$ in the second column
of Table~\ref{tab:Ktilde}, with the same overall factor removed as in
the real contribution---see Eq.~\ref{eq:IntKG}---for ease of comparison.

It can be seen from Table~\ref{tab:Ktilde} that the divergent parts of
$\widetilde{K}$ and $K$ cancel, leaving behind a finite remainder.
Contributions such as $(2,2)$ and $(5,5)$ from the square of matrix
elements $R2$ and its $u$-channel counterpart $R5$ as well as the $(2,5)$
crossed $tu$-channel contribution must necessarily have no divergent
terms as {\em the virtual counterpart does not exist}. (The same is true
for the virtual diagrams (4), (5) in Fig.~\ref{fig:nlo} which have no
real-photon counterparts.) This is indeed seen to hold.

Additionally, we have dropped terms that are collinearly divergent as
the fermion mass goes to zero; we will discuss these terms in the next
section. For example, it appears as though there is no soft divergent
contribution from the $(1,3)$ term, as can be seen from the relevant entry
in Table~\ref{tab:Ktilde}; however, there is indeed a soft divergent
contribution which is also collinear divergent. We will deal with such
terms in the next section. In fact, collinear divergent terms arise in
all contributions: the {\em thermal photon}, {\em thermal fermion} and
the {\em thermal anti-fermion} terms. {\em All} soft IR divergent terms
are correctly factored into the $\widetilde{K}$ contributions for the
real terms; hence the soft IR divergent terms which are also collinear
divergent are contained within the $\widetilde{K}$ contributions, so
that the $\widetilde{G}$ contributions are IR finite. However, some
of the {\em soft IR finite terms} which are collinear divergent are seen
to be partially contained in the $\widetilde{K}$ and partially in the
$\widetilde{G}$ contributions so that the GY separation for collinear
divergences is imperfect. The same is true for the virtual $K$ and $G$
contributions as well. However, overall the collinear divergences do
cancel between the real and virtual contributions. We will show the
cancellation of these terms in the next section.

\subsection{Cancellation of collinear divergences}
\label{ssec:collinear}

As mentioned earlier, it turns out that both the $\widetilde{G}$
and $\widetilde{K}$ contributions contain collinear divergences,
the sum of which cancel against the sum of the $G$ and $K$ collinear
divergent contributions. It is therefore more convenient to discuss
the collinear divergences arising from the sum of $\widetilde{G}$ and
$\widetilde{K}$ (and from the sum of $K$ and $G$ from the virtual photon
contribution). There are three sets of contributing terms, arising
from the case when the photon is thermal (with both $\widetilde{G}$
and $\widetilde{K}$ contributing), the fermion is thermal, and the
anti-fermion is thermal. Recall that when the photon is thermal, the
photon contributes via the thermal part of its propagator in the virtual
case, while it contributes via the thermal part of its phase space in
the real photon case, and similarly for the other two cases.

In the case of real photons, we have the factor $n_B(\omega)$ in
the phase space for thermal photons, while we have $(-n_F(\omega_t))$ for
thermal fermions (anti-fermions); see Eq.~\ref{eq:phase}. Hence we
compute the collinear contributions in $\sigma^{\gamma,
Real}_{i,j}$, the {\em thermal photon} contribution to the cross
section, as given in Eq.~\ref{eq:IntKG}, as well as from the {\em
thermal fermion} contributions, given by
\begin{align}
\sigma_{i,j}^{f, Real}
& = -\frac{e^2 \pi \vert \lambda\vert^4}{32 \sqrt{s} Q}
\frac{1}{2(2\pi)^3} \int
{\rm d} \omega_t ~ n_F(\omega_t) ~ Int^{f}_{i,j}~,
\label{eq:Intf}
\end{align}
with an analogous definition for the {\em thermal anti-fermion}
contribution, $\sigma_{i,j}^{\bar{f}, Real}$. These contributions will
cancel against the collinear divergent terms from the corresponding
virtual contributions. Note that in the case of virtual contributions,
the {\em thermal photon}, {\em thermal fermion} and {\em thermal
anti-fermion} contributions come from the corresponding thermal parts
of their propagators in Diagrams (1)--(7) in Figs.~\ref{fig:nlo} and
\ref{fig:nlo_self}. For more details on the virtual contribution, see
Appendix~\ref{sec:virtual} or Ref.~\cite{Butola:2024oia}.

The relevant real and virtual collinear contributions are listed in
Tables~\ref{tab:coll_gamma}, \ref{tab:coll_f} and \ref{tab:coll_fbar},
where only the leading $s$-wave results with $v \to 0$ are shown;
the exact expressions are available on-line as a Mathematica notebook
\cite{onlinereal}.

It may be noted that the logarithms are not precisely the same in the
real and virtual case; they merely match in the collinear (massless)
limit. In the case of {\em thermal photons}, the structure of the
logarithms that contribute in the virtual ($V$) and real ($R$) case is
given by
\begin{align}
L_V & = \log\left[\frac{H-P}{H+P}\right] \approx
 \log\left[\frac{m_f^2 }{4H^2 - m_f^2}\right]~, \nonumber \\
L_R & = \frac{1}{2} \left(\log\left[\frac{m_f^2}
	{4H^2 - 4 H \omega - m_f^2}\right] +
	\log\left[\frac{m_f^2}{4H^2 + 4 H \omega - m_f^2}
	\right] \right)~, \nonumber \\
L'_R & = \frac{1}{2} \left(\log\left[\frac{m_f^2}
	{4H^2 - 4 H \omega - m_f^2}\right] -
	\log\left[\frac{m_f^2}{4H^2 + 4 H \omega - m_f^2}
	\right] \right)~.
\label{eq:collgamma}
\end{align}
Since $\omega$ is strictly less than $(s - m_f^2)/(2\sqrt{s}) = H -
m_f^2/(4H)$, we can expand $L_R$ and $L'_R$ as
\begin{align}
L_R & = \frac{1}{2} \left(\log\left[\frac{m_f^2}
	{4H^2 - m_f^2}\left( 1 + \frac{4 H \omega}
	{4H^2 - m_f^2}\right) \right] +
	\log\left[\frac{m_f^2} {4H^2 - m_f^2}
	\left( 1 - \frac{4 H \omega}
	{4H^2 - m_f^2}\right) \right] \right),
	\nonumber \\
	& \approx \frac{1}{2} \left[2 \log\left[\frac{m_f^2}
	{4H^2 - m_f^2}\right] + {\cal{O}}(\omega)^2 \right]
	\equiv L_V + {\cal{O}}(\omega)^2~, \nonumber \\
L'_R & = \frac{1}{2} \left(\log\left[\frac{m_f^2}
	{4H^2 - m_f^2}\left( 1 + \frac{4 H \omega}
	{4H^2 - m_f^2}\right) \right] -
	\log\left[\frac{m_f^2} {4H^2 - m_f^2}
	\left( 1 - \frac{4 H \omega}
	{4H^2 - m_f^2}\right) \right] \right),
	\nonumber \\
	& \approx \frac{1}{2} \left[2 \left[\frac{4 H \omega}
	{4H^2 - m_f^2}\right] + {\cal{O}}(\omega)^2 \right]~,
\label{eq:collr}
\end{align}
where we have expanded the log terms appropriately. We have retained terms
only till order ${\cal{O}}(\omega)$ since higher order terms contribute
at order ${\cal{O}}((T/m_\chi)^4)$ or higher, and are hence small. It
turns out that the coefficient of $L'_R$, when summed over all $(i,j)$
as shown in Eqs.~\ref{eq:sigma_nlo_real}, \ref{eq:sigma_m2}, equals
$(0 \times m_\chi^2 + {\cal{O}}(m_f^2))$ and hence this contribution
vanishes in the collinear limit as $m_f \to 0$. The $L_R$ contributions
remain and are listed in Table~\ref{tab:coll_gamma}.

\begin{table}[bhp]
\centering
\begin{tabular}{|c|c|c|} \hline
\multicolumn{3}{|c|}{\bf Thermal Photon Collinear Contribution
\rule{0pt}{16pt}} \\ \hline
$i,j$ \rule{0pt}{16pt}
& Real & Virtual \\ \hline
$1,1$ \rule{0pt}{16pt}
& $-\omega \frac{48 m_\chi^2}{3 m_\phi^4} L_R$ & 
$\omega \frac{48 m_\chi^2}{3 m_\phi^4} L_V$ 
\\ \hline
$2,2$ \rule{0pt}{16pt}
& 0 &
--
\\ \hline
$3,3$ \rule{0pt}{16pt}
& $-\omega \frac{48 m_\chi^2}{3 m_\phi^4} L_R$ & 
$\omega \frac{48 m_\chi^2}{3 m_\phi^4} L_V$
\\ \hline
$4,4$ \rule{0pt}{16pt}
& $-\omega \frac{16 m_\chi^2}{m_\phi^4} L_R$ &
$\omega \frac{16 m_\chi^2}{m_\phi^4} L_V$ 
\\ \hline
$5,5$ \rule{0pt}{16pt}
& 0 &
--
\\ \hline
$6,6$ \rule{0pt}{16pt}
& $-\omega \frac{48 m_\chi^2}{3 m_\phi^4} L_R$ & 
$\omega \frac{48 m_\chi^2}{3 m_\phi^4} L_V$ 
\\ \hline
$1,2$ \rule{0pt}{16pt}
& $ - \omega \frac{32}{3 m_\phi^6} (6 m_\chi^4 - 3 m_\chi^2 m_f^2 ) L_R$ &
$ \omega \frac{32}{3 m_\phi^6} (6 m_\chi^4 - 3 m_\chi^2 m_f^2 ) L_V$
\\ \hline
$1,3$ \rule{0pt}{16pt}
& $-\frac{16 (4 m_\chi^4 - 2 m_\chi^2 m_f^2)}{\omega m_\phi^4} L_R -
\omega \frac{16 m_f^2}{m_\phi^4} L_R$ & 
$\frac{16 (4 m_\chi^4 - 2 m_\chi^2 m_f^2)}{\omega m_\phi^4}
L_V+
\omega \frac{16 m_f^2}{m_\phi^4} L_V$ 
\\ \hline
$1,4$ \rule{0pt}{16pt}
& $-\omega \frac{32 m_\chi^2}{m_\phi^4} L_R$ &
$\omega \frac{32 m_\chi^2}{m_\phi^4} L_V$
\\ \hline
$1,5 + 2,4$ \rule{0pt}{16pt}
& $-\omega\frac{32 m_\chi^2(4 m_\chi^2 - 3 m_f^2)}{m_\phi^6} L_R$ &
$\omega\frac{32 m_\chi^2(4 m_\chi^2 - 3 m_f^2)}{m_\phi^6} L_V$
\\ \hline
$1,6 + 3,4$ \rule{0pt}{16pt}
& $-\frac{32 (-2 m_\chi^2 + m_f^2)^2}{\omega m_\phi^4} L_R $ &
$\frac{32 (-2 m_\chi^2 + m_f^2)^2}{\omega m_\phi^4} L_V $ 
\\ \hline
$2,3$ \rule{0pt}{16pt}
& $-\omega \frac{32 m_\chi^2(2 m_\chi^2 - m_f^2)}{m_\phi^6} L_R$ &
$\omega \frac{32 m_\chi^2(2 m_\chi^2 - m_f^2)}{m_\phi^6} L_V$
\\ \hline
$2,5$ \rule{0pt}{16pt}
& 0 &
--
\\ \hline
$2,6 + 3,5$ \rule{0pt}{16pt}
& $-\omega\frac{32 m_\chi^2(4 m_\chi^2-3 m_\chi^2)} {m_\phi^6} L_R $ &
$\omega\frac{32 m_\chi^2(4 m_\chi^2-3 m_\chi^2)} {m_\phi^6} L_V$
\\ \hline
$3,6$ \rule{0pt}{16pt}
& $-\omega\frac{32 m_\chi^2} {m_\phi^4} L_R $ &
$\omega\frac{32 m_\chi^2} {m_\phi^4} L_V$
\\ \hline
$4,5$ \rule{0pt}{16pt}
& $-32\omega m_\chi^2\frac{2 m_\chi^2 - m_f^2)}{m_\phi^6} L_R$ &
$32\omega m_\chi^2\frac{2 m_\chi^2 - m_f^2)}{m_\phi^6} L_V$
\\ \hline
$4,6$ \rule{0pt}{16pt}
& $-\frac{16 (4 m_\chi^4 - 2 m_\chi^2 m_f^2)}{\omega m_\phi^4}
L_R - \omega \frac{16 m_f^2}{m_\phi^4} L_R$ & 
$\frac{16 (4 m_\chi^4 - 2 m_\chi^2 m_f^2)}{\omega m_\phi^4} L_V
+ \omega \frac{16 m_f^2}{m_\phi^4} L_V$
\\ \hline
$5,6$ \rule{0pt}{16pt}
& $-32\omega m_\chi^2\frac{2 m_\chi^2 - m_f^2)}{m_\phi^6} L_R$ &
$32\omega m_\chi^2\frac{2 m_\chi^2 - m_f^2)}{m_\phi^6} L_V$
\\ \hline
\end{tabular}
\caption{\small \em Collinear divergences in various terms of
the real and virtual {\em thermal photon} contributions.
Overall factors as in Eq.~\ref{eq:IntKG} have
been removed from the terms listed here. Divergent terms from real and
virtual contributions cancel. See text for the definition of
log terms $L_R$ and $L_V$.}
\label{tab:coll_gamma}
\end{table}

It can be seen from Table~\ref{tab:coll_gamma} that, upon using
the definitions of $L_V$ and $L_R$ in Eqs.~\ref{eq:collgamma} and
\ref{eq:collr}, the collinear divergences (containing both soft IR
divergent and IR finite terms) cancel between the real and virtual photon
contributions when the photon is thermal. Note that terms proportional
to $m_f^2$ in the coefficient of the log terms vanish in the collinear
limit and hence are not collinear divergent.

A similar analysis can be done when the fermion (anti-fermion) is
thermal. The results are tabulated in Tables~\ref{tab:coll_f}
(\ref{tab:coll_fbar}). Here 
we have the log terms $L_{R_f}$ and $L'_{R_f}$ for the real {\em thermal
fermion} contribution:
\begin{align}
L_{R_f} & = \frac{1}{2} \left(
\log\left[
\frac{2H(\omega_t-K_t)-m_f^2}{2H(\omega_t+K_t)-m_f^2}\right]
+ \log\left[
\frac{2H(\omega_t-K_t)+m_f^2}{2H(\omega_t+K_t)+m_f^2}\right]
	\right)~, \nonumber \\
L'_{R_f} & = \frac{1}{2} \left(
\log\left[
\frac{2H(\omega_t-K_t)-m_f^2}{2H(\omega_t+K_t)-m_f^2}\right]
- \log\left[
\frac{2H(\omega_t-K_t)+m_f^2}{2H(\omega_t+K_t)+m_f^2}\right]
	\right)~,
\label{eq:coll_rf}
\end{align}

\begin{table}[bhp]
\centering
\begin{tabular}{|c|c|c|} \hline
\multicolumn{3}{|c|}{\bf Thermal Fermion Collinear Contribution
\rule{0pt}{16pt}} \\ \hline
$i,j$ \rule{0pt}{16pt}
& Real & Virtual \\ \hline
$1,1$ \rule{0pt}{16pt}
& $-\omega_t \frac{16 m_\chi^2}{m_\phi^4} L_{R_f}$ & 
$\omega_t \frac{16 m_\chi^2}{m_\phi^4} L_{V_f}$ 
\\ \hline
$2,2$ \rule{0pt}{16pt}
& 0 &
--
\\ \hline
$3,3$ \rule{0pt}{16pt}
& 0 &
0
\\ \hline
$4,4$ \rule{0pt}{16pt}
& $-\omega_t \frac{16 m_\chi^2}{m_\phi^4} L_{R_f}$ & 
$\omega_t \frac{16 m_\chi^2}{m_\phi^4} L_{V_f}$ 
\\ \hline
$5,5$ \rule{0pt}{16pt}
& 0 &
--
\\ \hline
$6,6$ \rule{0pt}{16pt}
& 0 &
0
\\ \hline
$1,2$ \rule{0pt}{16pt}
& $ -\omega_t \frac{32 m_\chi^2 (2 m_\chi^2 - m_f^2 )} {m_\phi^6} L_{R_f}$ &
$ \omega_t \frac{32 m_\chi^2 (2 m_\chi^2 - m_f^2 )} {m_\phi^6} L_{V_f}$
\\ \hline
$1,3$ \rule{0pt}{16pt}
& $ \omega_t \frac{8 (4 m_\chi^2 - 3 m_f^2 )} {m_\phi^4} L_{R_f}$ &
$ -\omega_t \frac{8 (4 m_\chi^2 - 3 m_f^2 )} {m_\phi^4} L_{V_f}$
\\ \hline
$1,4$ \rule{0pt}{16pt}
& $ -\omega_t \frac{32 m_\chi^2} {m_\phi^4} L_{R_f}$ &
$ \omega_t \frac{32 m_\chi^2} {m_\phi^4} L_{V_f}$
\\ \hline
$1,5 + 2,4$ \rule{0pt}{16pt}
& $ -\omega_t \frac{32 m_\chi^2 (4 m_\chi^2 - 3 m_f^2 )} {m_\phi^6}
L_{R_f}$ &
$ \omega_t \frac{32 m_\chi^2 (4 m_\chi^2 - 3 m_f^2 )} {m_\phi^6}
L_{V_f}$ 
\\ \hline
$1,6 + 3,4$ \rule{0pt}{16pt}
& $ \omega_t \frac{16 (2 m_\chi^2 - m_f^2)^2} {m_\chi^2 m_\phi^4} L_{R_f}$ &
$ -\omega_t \frac{16 (2 m_\chi^2 - m_f^2)^2} {m_\chi^2 m_\phi^4} L_{V_f}$
\\ \hline
$2,3$ \rule{0pt}{16pt}
& 0 &
0
\\ \hline
$2,5$ \rule{0pt}{16pt}
& 0 &
--
\\ \hline
$2,6 + 3,5$ \rule{0pt}{16pt}
& 0 &
0
\\ \hline
$3,6$ \rule{0pt}{16pt}
& 0 & 
 0
\\ \hline
$4,5$ \rule{0pt}{16pt}
& $ -\omega_t \frac{32 m_\chi^2 (2 m_\chi^2 - m_f^2 )} {m_\phi^6} L_{R_f}$ &
$ \omega_t \frac{32 m_\chi^2 (2 m_\chi^2 - m_f^2 )} {m_\phi^6} L_{V_f}$
\\ \hline
$4,6$ \rule{0pt}{16pt}
& $ \omega_t \frac{8 m_\chi^2 (4 m_\chi^2 - 3 m_f^2 )} {m_\phi^4}
L_{R_f}$ &
$ -\omega_t \frac{8 m_\chi^2 (4 m_\chi^2 - 3 m_f^2 )} {m_\phi^4}
L_{V_f}$
\\ \hline
$5,6$ \rule{0pt}{16pt}
& 0 &
0
\\ \hline
\end{tabular}
\caption{\small \em As in Table~\ref{tab:coll_gamma} for collinear
divergences in various terms of real and virtual contributions
from {\em thermal fermions}. See text for a definition of the log
terms $L_{R_f}$ and $L_{V_f}$.}
\label{tab:coll_f}
\end{table}

For the virtual {\em thermal fermion} contribution we have the log terms $L_{V_f}$ and $L'_{V_f}$, where
\begin{align}
L_{V_f} & = \frac{1}{2} \left(
\log\left[ \frac{H \omega_t - K_t P' - m_f^2} {H \omega_t + K_t P' -
m_f^2} \right] + \log\left[
\frac{H \omega_t - K_t P' + m_f^2} {H \omega_t + K_t P' +
m_f^2} \right] \right)~, \nonumber \\
L'_{V_f} & = \frac{1}{2} \left(
\log\left[ \frac{H \omega_t - K_t P' - m_f^2} {H \omega_t + K_t P' -
m_f^2} \right] 
- \log\left[ \frac{H \omega_t - K_t P' + m_f^2} {H \omega_t + K_t P' +
m_f^2} \right] \right)~.
\label{eq:coll_vf}
\end{align}

\begin{table}[bhp]
\centering
\begin{tabular}{|c|c|c|} \hline
\multicolumn{3}{|c|}{\bf Thermal Anti-Fermion Collinear Contribution
\rule{0pt}{16pt}} \\ \hline
$i,j$ \rule{0pt}{16pt}
& Real & Virtual \\ \hline
$1,1$ \rule{0pt}{16pt}
& 0 
 & 0
\\ \hline
$2,2$ \rule{0pt}{16pt}
& 0 &
--
\\ \hline
$3,3$ \rule{0pt}{16pt}
& $-\omega_t \frac{16 m_\chi^2}{m_\phi^4} L_{R_f}$ & 
$\omega_t \frac{16 m_\chi^2}{m_\phi^4} L_{V_f}$
\\ \hline
$4,4$ \rule{0pt}{16pt}
& 0 &
0
\\ \hline
$5,5$ \rule{0pt}{16pt}
& 0 &
--
\\ \hline
$6,6$ \rule{0pt}{16pt}
& $-\omega_t \frac{16 m_\chi^2}{m_\phi^4} L_{R_f}$ & 
$\omega_t \frac{16 m_\chi^2}{m_\phi^4} L_{V_f}$
\\ \hline
$1,2$ \rule{0pt}{16pt}
& 0 &
0
\\ \hline
$1,3$ \rule{0pt}{16pt}
& $ \omega_t \frac{8 (4 m_\chi^2 - 3 m_f^2 )} {m_\phi^4} L_{R_f}$ &
$ -\omega_t \frac{8 (4 m_\chi^2 - 3 m_f^2 )} {m_\phi^4} L_{V_f}$
\\ \hline
$1,4$ \rule{0pt}{16pt}
& 0 &
0
\\ \hline
$1,5 + 2,4$ \rule{0pt}{16pt}
& 0 &
0
\\ \hline
$1,6 + 3,4$ \rule{0pt}{16pt}
& $ \omega_t \frac{16 (2 m_\chi^2 - m_f^2)^2} {m_\chi^2 m_\phi^4} L_{R_f}$ &
$ -\omega_t \frac{16 (2 m_\chi^2 - m_f^2)^2} {m_\chi^2 m_\phi^4} L_{V_f}$
\\ \hline
$2,3$ \rule{0pt}{16pt}
& $ -\omega_t \frac{32 (2 m_\chi^2 - m_f^2)} {m_\phi^6} L_{R_f}$ &
$ \omega_t \frac{32 (2 m_\chi^2 - m_f^2)} {m_\phi^6} L_{V_f}$
\\ \hline
$2,5$ \rule{0pt}{16pt}
& 0 &
--
\\ \hline
$2,6 + 3,5$ \rule{0pt}{16pt}
& $ -\omega_t \frac{32 m_\chi^2(4 m_\chi^2 - 3 m_f^2)} {m_\phi^6} L_{R_f}$ &
$ \omega_t \frac{32 m_\chi^2(4 m_\chi^2 - 3 m_f^2)} {m_\phi^6} L_{V_f}$
\\ \hline
$3,6$ \rule{0pt}{16pt}
& $-\omega_t \frac{32 m_\chi^2}{m_\phi^4} L_{R_f}$ & 
$\omega_t \frac{32 m_\chi^2}{m_\phi^4} L_{V_f}$
\\ \hline
$4,5$ \rule{0pt}{16pt}
& 0 &
0
\\ \hline
$4,6$ \rule{0pt}{16pt}
& $ \omega_t \frac{8 (4 m_\chi^2 - 3 m_f^2 )} {m_\phi^4} L_{R_f}$ &
$ -\omega_t \frac{8 (4 m_\chi^2 - 3 m_f^2 )} {m_\phi^4} L_{V_f}$
\\ \hline
$5,6$ \rule{0pt}{16pt}
& $ -\omega_t \frac{32 m_\chi^2 (2 m_\chi^2 - m_f^2 )} {m_\phi^6} L_{R_f}$ &
$ \omega_t \frac{32 m_\chi^2 (2 m_\chi^2 - m_f^2 )} {m_\phi^6} L_{V_f}$
\\ \hline
\end{tabular}
\caption{\small \em As in Table~\ref{tab:coll_gamma} for collinear
divergences in various terms of real and virtual contributions
from {\em thermal anti-fermions}. See text for a definition of the log
terms $L_{R_f}$ and $L_{V_f}$.}
\label{tab:coll_fbar}
\end{table}

We find that the coefficient of every contribution to $L'_{V_f}$ is
proportional to $m_f^2$ and hence this contribution vanishes in the
collinear limit. As with $L'_R$, the sum over all $(i,j)$ of $L'_{R_f}$
yields a coefficient which has the form $(0 \times m_\chi^2 +
{\cal{O}}(m_f^2))$, so that these contributions also vanish in the
collinear limit. We are left with terms proportional to $L_{V_f}$ and
$L_{R_f}$, as listed in Tables~\ref{tab:coll_f} and \ref{tab:coll_fbar}.
On replacing $P' = \sqrt{H^2 - m_f^2}$, expanding and
retaining the leading terms, we can write these terms as
\begin{align}
L_{R_f} & \approx \frac{1}{2} \left(2 \log\left[
\frac{\omega_t - K_t}{\omega_t + K_t} \right] \right)
\approx \log\left[ \frac{m_f^2}{4 \omega_t^2 - m_f^2} \right]~, \nonumber \\
L_{V_f} & \approx \frac{1}{2} \left(2 \log\left[
\frac{\omega_t - K_t}{\omega_t + K_t} \right] 
\right) = L_{R_f}~.
\label{eq:collf}
\end{align}
Notice the unusual structure of the log terms, both {\em independent} of
the external energy $H$ (except indirectly through the limits on
$\omega_t$). To our knowledge, such a dependence has been observed for
the first time. With the identification of $L_{V_f} = L_{R_f}$, again, we
see from Tables~\ref{tab:coll_f}, \ref{tab:coll_fbar} that the collinear
divergent terms arising from thermal fermion and thermal anti-fermion
pieces also cancel between the real and virtual terms. Hence the entire
cross section is both soft IR finite and free of collinear divergences
to ${\cal{O}}(\alpha T^2)$.

It may be useful to point out that the structure of the log terms
$L_R$, $L_{V_f}$ and $L_{R_f}$ is non-trivial and furthermore, $\omega$ (or
$\omega_t$) dependent and hence can only be exactly numerically computed.
Our approximations in Eqs.~\ref{eq:collr} and \ref{eq:collf} allow for
the cancellations to be demonstrated at the level of the integrands
themselves, without having to perform the $\omega$ ($\omega_t$)
integrations. While it is satisfying to see the cancellation of these
collinear divergences, it must be kept in mind that heavy fermions such
as $\tau$ leptons or $b$ quarks can significantly contribute through
these logs, which may substantially alter our results shown in the next
section for the nearly massless case. Again, the exact expressions are
available on-line \cite{onlinereal} for use in numerical computations.

We now discuss the finite remainder which is the thermal contribution to
the real photon cross section.

\subsection{Finite remainder and the thermal real photon cross section}

The total {\em thermal photon} contribution to the real photon dark matter
annihilation process arises from the sum of the finite $\widetilde{G}$
and the finite $Int_{i,j}^{\widetilde{K}_{fin}} = (\widetilde{K} +
K_{div})$; we have already discussed the latter in the previous section
and the corresponding results are listed in Table~\ref{tab:Ktilde}. The
contribution $Int_{i,j}^{\widetilde{G}}$ from the {\em thermal photon}
$\widetilde{G}$ part from various $(i,j)$ terms (see Eq.~\ref{eq:IntKG})
is shown in Table~\ref{tab:Gffbartilde}, modulo the collinear divergences,
which have been shown to cancel in Section~\ref{ssec:collinear}.

\begin{table}[htp]
\centering
\begin{tabular}{|c|c|c|c|}
\hline
$i,j$ \rule{0pt}{16pt}
& {\bf Real} $Int^{\widetilde{G}}_{i,j}$
from $Int^\gamma_{i,j}$ & {\bf Real} $Int^f_{i,j}$
& {\bf Real} $Int^{\overline{f}}_{i,j}$ \\ \hline
$1,1$ \rule{0pt}{16pt}
& $\omega (\frac{32 m_\chi^2}{m_\phi^4})$
& $-K_t (\frac{32 m_\chi^2 - 12 m_f^2}{m_\phi^4})$
 & 
 0
\\ \hline
$2,2$ \rule{0pt}{16pt}
& $\omega m_\chi^4(\frac{256 m_\chi^2 - 96 m_f^2}{m_\phi^8} )$
& 0 & 
0
\\ \hline
$3,3$ \rule{0pt}{16pt}
& $\omega (\frac{32 m_\chi^2}{m_\phi^4})$ 
 &
0
 & 
$-K_t (\frac{32 m_\chi^2 - 12 m_f^2}{m_\phi^4})$
\\ \hline
$4,4$ \rule{0pt}{16pt}
& $\omega (\frac{32 m_\chi^2}{m_\phi^4})$ 
& $-K_t (\frac{32 m_\chi^2 - 12 m_f^2}{m_\phi^4})$ 
 &
 0
\\ \hline
$5,5$ \rule{0pt}{16pt}
& $\omega m_\chi^4(\frac{256 m_\chi^2 - 96 m_f^2}{m_\phi^8})$
& 0 & 
0
\\ \hline
$6,6$ \rule{0pt}{16pt}
& $\omega ( \frac{32 m_\chi^2}{m_\phi^4})$ 
 &
0
 & 
$-K_t (\frac{32 m_\chi^2 - 12 m_f^2}{m_\phi^4})$ 
\\ \hline
$1,2$ \rule{0pt}{16pt}
 & 
$\omega m_\chi^2 (\frac{64 m_\chi^2 - 32 m_f^2}{m_\phi^6})$
&
$K_t m_\chi^2 (\frac{64 m_\chi^2}{m_\phi^6})$ 
& 
0 
\\ \hline
$1,3$ \rule{0pt}{16pt}
 & $-\omega ( \frac{64 m_\chi^2 - 32 m_f^2}{m_\phi^4})$ 
 &
$-K_t (\frac{32 m_\chi^2}{m_\phi^4})$ 
&
$-K_t (\frac{32 m_\chi^2}{m_\phi^4})$ 
\\ \hline
$1,4$ \rule{0pt}{16pt}
&
$\omega (\frac{32 m_\chi^2 + 16 m_f^2}{m_\phi^4})$
 &
 $-K_t ( \frac{64 m_\chi^2 - 32 m_f^2}{m_\phi^4})$ 
 &
 $-K_t ( \frac{8 m_f^2}{m_\phi^4})$ 
\\ \hline
$1,5+ 2,4$ \rule{0pt}{16pt}
 & 
$-\omega m_\chi^2 ( \frac{128 m_\chi^2 + 32 m_f^2}{m_\phi^6} )$ 
& 
$K_t m_\chi^2(\frac{128 m_\chi^2 - 160 m_f^2}{m_\phi^6})$
& 
$K_t m_\chi^2(\frac{32 m_f^2}{m_\phi^6})$
\\ \hline
$1,6+3,4$ \rule{0pt}{16pt}
 & 0
 &
$ -K_t (\frac{64 m_\chi^2 - 32 m_f^2}{m_\phi^4})$
 &
$ -K_t (\frac{64 m_\chi^2 - 32 m_f^2}{m_\phi^4})$
\\ \hline
$2,3$ \rule{0pt}{16pt}
 & $-\omega m_\chi^2(\frac{192 m_\chi^2}{m_\phi^6} )$ 
 &
 0
 &
$K_t m_\chi^2(\frac{64 m_\chi^2}{m_\phi^6})$
\\ \hline
$2,5$ \rule{0pt}{16pt}
 & $\omega m_\chi^4(\frac{256 m_\chi^2 - 256 m_f^2}{m_\phi^8} )$ 
 &
$ -K_t m_\chi^2 (\frac{64 m_\chi^2}{3m_\phi^8})$
 & 
$ -K_t m_\chi^2 (\frac{64 m_\chi^2}{3m_\phi^8})$
\\ \hline
$2,6+3,5$ \rule{0pt}{16pt}
 & $-\omega m_\chi^2(\frac{128 m_\chi^2 + 32 m_f^2}{m_\phi^6} )$ 
 &
$ K_t m_\chi^2(\frac{32 m_f^2}{m_\phi^6})$
 & 
$ K_t m_\chi^2(\frac{128 m_\chi^2 - 160 m_f^2}{m_\phi^6})$
\\ \hline
$3,6$ \rule{0pt}{16pt}
 & $ \omega ( \frac{32 m_\chi^2 + 16 m_f^2}{m_\phi^4})$ 
 &
$ -K_t (\frac{8 m_f^2}{m_\phi^4})$
 &
$ -K_t (\frac{64 m_\chi^2 - 32 m_f^2}{m_\phi^4})$
\\ \hline
$4,5$ \rule{0pt}{16pt}
 & $-\omega m_\chi^2(\frac{192 m_\chi^2}{m_\phi^6} )$ 
 &
$ K_t m_\chi^2 (\frac{64 m_\chi^2}{m_\phi^6})$
 & 
 0
\\ \hline
$4,6$ \rule{0pt}{16pt}
 & $-\omega ( \frac{64 m_\chi^2 - 32 m_f^2}{m_\phi^4})$ 
 &
$ -K_t (\frac{32 m_\chi^2}{m_\phi^4})$
 &
$ -K_t (\frac{32 m_\chi^2}{m_\phi^4})$
\\ \hline
$5,6$ \rule{0pt}{16pt}
& 
$\omega m_\chi^2 ( \frac{64 m_\chi^2 - 32 m_f^2}{m_\phi^6})$ 
& 0
&
$ K_t m_\chi^2 (\frac{64 m_\chi^2}{m_\phi^6})$
\\ \hline
\end{tabular}
\caption{\small \em The finite real photon contributions from various
diagrams: the {\em thermal photon} $\widetilde{G}$ contribution
(Eq.~\ref{eq:IntKG}) $Int^{\widetilde{G}}_{i,j}$, and the {\em thermal
fermion and anti-fermion} contributions $Int^f_{i,j}$ (Eq.~\ref{eq:Intf})
and $Int^{\overline{f}}_{i,j}$ contributions. The total NLO thermal
cross section is given by the sum of these contributions with the
finite $\widetilde{K}$ combination, $Int^{\widetilde{K}_{fin}}_{i,j}$
(see Table~\ref{tab:Ktilde}).}
\label{tab:Gffbartilde}
\end{table}

The thermal fermion (and anti-fermion) contributions $Int_{i,j}^f$ and
$Int_{i,j}^{\overline{f}}$ (see Eq.~\ref{eq:Intf}) are IR finite and can
be simply added to the thermal photon contribution. These are also listed
(again, modulo the collinear divergent terms which were shown to cancel
in Section~\ref{ssec:collinear}) in Table~\ref{tab:Gffbartilde}. The
total NLO thermal cross section is thus given by the sum of all these
terms. The detailed results are again listed online \cite{onlinereal}.
Here we present only the $s$-wave results in the limit that the dark
matter momentum $Q$ is small. The total NLO thermal cross section,
defined in Eq.~\ref{eq:sigma_gffbar} for the process $\chi \chi \to f
\overline{f} (\gamma)$ is then given by
\begin{align}
\sigma_{NLO}^{Real} & = \frac{\alpha \vert \lambda\vert^4}{128 v_{rel}}
 \, \frac{T^2}{m_\chi^2} ~m_f^2 \left[ \frac{16}{3 m_\phi^4} +
 \frac{160 m_\chi^4}{9 m_\phi^8} \right]~,
\label{eq:sigmaNLOReal}
\end{align}
where again we have retained terms to order ${\cal{O}}(T^2, m_f^2)$.
It can be seen that the total real photon thermal cross section is
again proportional to the fermion mass squared, and hence is helicity
suppresssed, although the individual terms in Tables~\ref{tab:Ktilde},
\ref{tab:Gffbartilde} are all not so. Hence the addition of real
photon emission/absorption to the leading order process $\chi \chi \to
f \overline{f}$ does not lift the helicity suppression, so that the
contribution is suppressed, just as was the case with the LO and the
virtual NLO result. Note that the ${\cal{O}}(1/m_\phi^6)$ term vanishes
to this order. Adding this to the virtual contribution that was computed
in Ref.~\cite{Butola:2024oia}, the total thermal contribution to the
dark matter annihilation cross section from $\chi \chi \to f \overline{f}$
as well as $\chi \chi \to f \overline{f} (\gamma)$ is given, to order
${\cal{O}}(\alpha T^2/m_\chi^2)$, by
\begin{align}
\sigma_{NLO}^{Real + Virtual} \, v_{rel} & = \frac{\alpha \vert
\lambda\vert^4}{128} \,
\frac{T^2}{m_\chi^2} ~ m_f^2 \left[-\frac{8}{3 m_\phi^4} + 
\frac{32 m_\chi^2}{3 m_\phi^6} +
\frac{544 m_\chi^4}{9 m_\phi^8} \right]~.
\label{eq:sigmaNLORealVirtual}
\end{align}
As mentioned earlier, the thermal average of this NLO cross section,
added to the LO term shown in Eq.~\ref{eq:sigma_LOv}, is the collision
term at order ${\cal{O}}(T^2/m_\chi^2)$, which determines how the DM relic
density evolved as the Universe cooled \cite{Gondolo:1990dk}. Both the
LO and NLO contributions are helicity suppressed. This is in contrast
to the case when the dark matter particles are Dirac type fermions,
which we briefly discuss below.

\subsubsection{Thermal real photon cross section for Dirac dark
matter}

As seen in Eq.~\ref{eq:sigma_LODirac}, the LO cross section when the
dark matter particles are of Dirac type is not helicity suppressed. In
this case, only the $t$-channel diagrams contribute to the cross
section. It turns out that the thermal cross section in this case is also
not helicity suppressed. Extracting only the $t$-channel terms,
we have, in the non-relativistic limit, the $s$-wave contribution,
\begin{align}
\sigma_{NLO}^{Real, Dirac} & = \frac{\alpha \vert \lambda\vert^4}{128 v_{rel}}
\, \frac{T^2}{m_\chi^2} \left[ \frac{16 m_\chi^2 + 5 m_f^2}{3 m_\phi^4}
+ \frac{32 m_\chi^2 ( m_\chi^2 - m_f^2)}{3 m_\phi^6}
+ \frac{32 m_\chi^4(2 m_\chi^2 - 3 m_f^2)}{3 m_\phi^8} \right]~.
\label{eq:sigmaNLORealDirac}
\end{align}
The total NLO cross section from both virtual and real photon
annihilation processes for Dirac dark matter is then
\begin{align}
\sigma_{NLO}^{Real + Virtual, Dirac} & =
\frac{\alpha \vert \lambda\vert^4}{128 v_{rel}}
\, \frac{T^2}{m_\chi^2} \left[-\frac{8 m_\chi^2 +7 m_f^2}{3 m_\phi^4}
+ \frac{16 m_\chi^2 ( 4 m_\chi^2 - 3 m_f^2)}{3 m_\phi^6} \right.
\nonumber \\
& \qquad \qquad \qquad \qquad \left. +
\frac{32 m_\chi^4(26 m_\chi^2 - 15 m_f^2)}{3 m_\phi^8} \right]~.
\label{eq:sigmaNLORealVirtualDirac}
\end{align}
It is interesting to note that the {\em ratio} of the NLO thermal
correction to the LO cross section for both Majorana and
Dirac dark matter particles in the $s$-wave limit is the same:
\begin{align}
\frac{\sigma_{NLO}}{\sigma_{LO}} & = -\frac{2\alpha \pi}{3} 
	\, \frac{T^2}{m_\chi^2}~,
\label{eq:ratio}
\end{align}
where we have considered the leading ${\cal{O}}(1/m_\phi^4)$ term in
both cases\footnote{Of course, the cross section {\em to} order
${\cal{O}}(\alpha)$ is the sum of the LO and NLO part; hence when we
refer to the ``NLO'' cross section here, we mean {\em only} the
${\cal{O}}(\alpha)$ contribution.}.

\section{Discussions and Conclusions}
\label{sec:concl}

In this paper, we have computed the thermal corrections to dark matter
annihilation processes with {\em real photon} emission/absorption, $\chi \chi
\to f \overline{f} (\gamma)$ via scalars, where $(\gamma)$ indicates
that the photon could be both emitted into, and absorbed from the heat
bath at temperature $T$. In an earlier paper ~\cite{Butola:2024oia}, we
had calculated the thermal corrections to the dark matter annihilation
process, $\chi \chi \to f \overline{f}$, {\em viz.}, the NLO {\em virtual
photon} contributions.

We have used the heavy scalar approximation to show explicitly in this
paper that the soft infra-red (IR) and collinear divergences cancel term
by term between the virtual and real contributions, leaving a finite
remainder; see Table~\ref{tab:Ktilde}. In both papers, the Grammer and
Yennie technique \cite{Grammer:1973db} was used to compute the IR finite
part of the cross sections. While this approach greatly simplifies the
calculation, however, it turns out that this technique only isolates
the soft IR divergences into the so-called $K$ ($\widetilde{K}$) photon
terms, but not the collinear divergences. Hence we have separately shown
the cancellation of the soft and collinear divergences between real
and virtual contributions, {\em term by term} for {\em thermal photon},
{\em thermal fermion}, and {\em thermal anti-fermion} contributions; see
Tables~\ref{tab:coll_gamma}, \ref{tab:coll_f}, and \ref{tab:coll_fbar}. In
particular, it turns out that the {\em structure} of the logarithmic
contributions in the case when the fermion (or anti-fermion) contributes
through its thermal parts (in the propagator or phase-space for virtual
and real cases respectively) is rather unusual; see Eq.~\ref{eq:collf}
and the cancellation of the collinear divergences non-trivial.

It was shown in Ref.~\cite{Beneke:2014gla} that the total NLO thermal
correction to the leading order Majorana dark matter annihilation cross
section, see Eq.~\ref{eq:sigma_LO}, computed to leading order in the
heavy scalar mass, ${\cal{O}}(1/m_\phi^4)$, vanishes in the limit of
zero fermion mass. In both our earlier work \cite{Butola:2024oia}
and this paper, we have calculated also the relatively suppressed
${\cal{O}}(1/m_\phi^6)$ and ${\cal{O}}(1/m_\phi^8)$ contributions,
apart from the leading ${\cal{O}}(1/m_\phi^4)$ result. The entire
thermal correction turns out to be helicity suppressed; see
Eq.~\ref{eq:sigmaNLORealVirtual} for the result to order
${\cal{O}}(m_f^2)$. On the other hand, neither the LO cross
section, Eq.~\ref{eq:sigma_LODirac}, nor the NLO thermal corrections
Eq.~\ref{eq:sigmaNLORealVirtualDirac}, are helicity suppressed in the case
of Dirac dark matter. Interestingly, it appears that, to leading order in
the heavy scalar mass, the thermal corrections may be insensitive to the
nature of the fermion-dark matter-scalar vertex since the calculation for
both Majorana and Dirac type dark matter particles yields the {\em same}
thermal NLO to LO cross section ratio, see Eq.~\ref{eq:ratio}. Of course,
the cross sections themselves are very different from one another, with
the Majorana annihilation process being helicity suppressed while the
Dirac one is not.

In Section~\ref{sec:real} we have presented only the leading $s$-wave
contributions, assuming that the dark matter is highly non-relativistic
(dark matter momentum $Q \to 0$). We have also shown only the leading
${\cal{O}}(\alpha\, T^2/m_\chi^2)$ thermal contribution, and retained terms to
order $m_f^2$ in the fermion masses. The complete expressions, without
these approximations, are available as a set of Mathematica notebooks
online \cite{onlinereal}. The thermal average of this annihilation
cross section (to be precise, $\langle \sigma v_{rel} \rangle_T$) is
the collision term in the Boltzmann equation that determines how the
dark matter number density varies in time; when the interaction rate
falls below the Hubble expansion rate of the Universe, the dark matter
freezes out into its presently observed distribution and determines the
dark matter relic density. Of course, if the dark matter production rate
is small to begin with, the dark matter density rather freezes-in to
its observed current value. Although the thermal contributions to the
collision term are seen to be small they may be significant especially
in freeze-in scenarios where $m_\chi/T \sim {\cal{O}}(1)$. In any
case, increasingly precise measurements of the dark matter relic
density \cite{ParticleDataGroup:2024cfk} indicate that more accurate
estimates of the dark matter annihilation cross section may be of use.
Hence thermal corrections to such cross sections may find their place
in such precision calculations.

\appendix
\renewcommand{\theequation}{\thesection.\arabic{equation}}
\setcounter{equation}{0}

\section{Thermal field theory}
\label{sec:tft}

\begin{figure}[htp]
\centering
\includegraphics[width=0.5\textwidth]{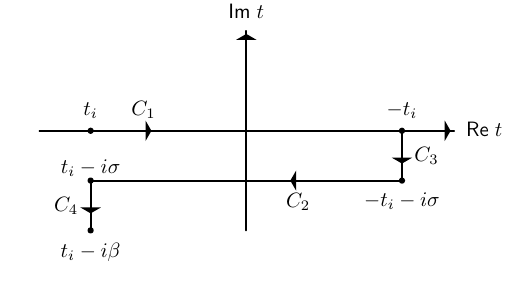}
\caption{\small \em The time path for real time formulation of thermal
field theories in the complex $t$ plane, where the $y$ axis corresponds
to $\Im t = \beta$, the inverse temperature.}
\label{fig:timepath}
\end{figure}

In the real-time formulation of thermal field theories \cite{Kobes:1985kc,
Niemi:1983ea, Rivers:1987hi, Altherr:1993tn}, the integration
in the complex time plane is defined over a special path shown in
Fig.~\ref{fig:timepath}, from an initial time, $t_i$ to a final time,
$t_i - i \beta$, where $\beta$ is the inverse temperature of the heat
bath, $\beta = 1/T$. The consequent field doubling---which can be of
type-1 (the physical fields; only these can occur on external legs),
or type-2 (the ``ghost'' fields)---leads to a $2 \times 2$ matrix form
for all propagators, with the off-diagonal elements of the propagator
allowing for conversion of one type into another.

All propagators can be written as the sum of two terms, one which is
{\em temperature-independent} and the other which contains the explicitly
thermal dependence which we call the {\em thermal} part.

\subsection{Feynman rules in thermal field theory}
\label{sec:feyn}

The scalar propagator is given by
\begin{equation}
i {{S}}_{\rm scalar}^{t_a, t_b} (p,m) = \left(
	\begin{array}{cc}
	\Delta(p) & 0 \\
	0 & \Delta^*(p) \end{array} \right) +
	2 \pi \delta(p^2-m^2) n_{\rm B}(\vert p^0 \vert) 
	\left(\begin{array}{cc}
	1 & e^{\vert p^0 \vert /(2T)} \\
	e^{\vert p^0 \vert /(2T)} & 1 
	\end{array} \right)~,
\label{eq:sprop}
\end{equation}
where $\Delta(p) = i/(p^2-m^2+i\epsilon)$, and $t_a, t_b~(=1,2)$ refer
to the field's thermal type. Only the {\em thermal} parts can convert
type-1 to type-2 fields, and vice versa; note that these contribute only
on mass-shell.

The photon propagator corresponding to a momentum $k$ is given in the
Feynman gauge by
\begin{align}
i {\cal D}^{t_a,t_b}_{\mu\nu} (k) =
-g_{\mu\nu} i {D}^{t_a,t_b} (k) =
- g_{\mu\nu} \, i {{S}}_{\rm scalar}^{t_a, t_b} (k,0)~,
\label{eq:thermalD}
\end{align}
while the fermion propagator at zero chemical potential is given by
\begin{align} \nonumber
i {\cal{S}}_{\rm fermion}^{t_a, t_b} (p,m) & = \left( \!\!\!
	\begin{array}{cc}
	S & 0 \\
	0 & S^*
	\end{array} \!\!\! \right)
	- 2 \pi S' \delta(p^2\!-\!m^2) n_{\rm F}(\vert p^0 \vert) 
	\left( \!\!\! \begin{array}{cc}
	1 & \!\!\!\!\!\epsilon(p_0) e^{\vert p^0 \vert /(2T)} \\
	-\epsilon(p_0) e^{\vert p^0 \vert /(2T)} & 1 
	\end{array} \!\!\! \right)~, \\
 & \equiv (\slashed{p}+m) \left(\begin{array}{cc}
 	F_p^{-1} & G_p^{-1} \\
	-G_p^{-1} & F_p^{*-1}
\end{array} \right)~, \nonumber \\
 & \equiv (\slashed{p}+m) \overline{S}^{t_a,t_b}(p,m)~,
\label{eq:fprop}
\end{align}
where $S = i/(\slashed{p} -m+i\epsilon)$, and $S' = (\slashed{p} +m)$.
The fermion propagator is proportional to $(\slashed{p}+m)$,
just as at $T=0$. The number operator for fermions and bosons is given
in Eq.~\ref{eq:nBF}.

\begin{figure}[htp]
\centering
\includegraphics[width=\textwidth]{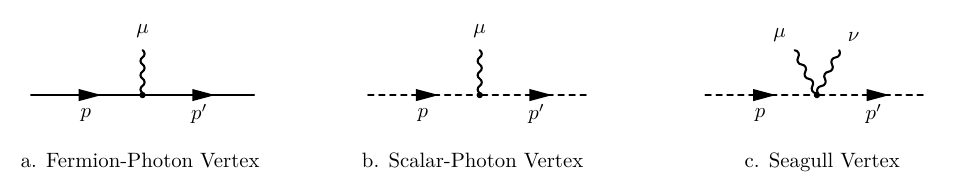}
\caption{\small \em Allowed vertices for fermion--photon and
scalar--photon interactions.}
\label{fig:feynman}
\end{figure}

The fermion--photon vertex factor is given by
$(-ie\gamma_\mu)(-1)^{t_\mu+1}$, where $t_\mu=1,2$ for the type-1 and
type-2 vertices. The scalar--photon vertex factor is $[-ie(p_\mu +
p'_\mu)](-1)^{t_\mu+1}$ where $p_\mu$ ($p'_\mu$) is the 4-momentum of
the scalar entering (leaving) the vertex, while the 2-scalar--2-photon
{\em seagull} vertex factor (see Fig.~\ref{fig:feynman}) is $[+2ie^2
g_{\mu\nu}](-1)^{t_\mu+1}$ (the factor `2' is dropped for a {\em tadpole}
vertex).

The DM-scalar-fermion vertex factor is $i \lambda P_L$; for details
on Feynman rules for Majorana particles at zero temperature, see
Ref.~\cite{Denner:1992vza}. An overall negative sign applies as usual
to the type-2 DM vertex; all fields at a vertex are of the same type,
and all external fields are of type-1.

\section{Some details of the virtual photon calculation}
\label{sec:virtual}

We present for completeness some details of the thermal NLO virtual
correction to $\chi \chi \to f \overline{f}$; for more details, see
Ref.~\cite{Butola:2024oia}. Apart from the contributions from Diagrams
1--5 in Fig.~\ref{fig:nlo} which are straightforward to compute, we also
have the contributions from Diagrams 6--7 in Fig.~\ref{fig:nlo_self}. The
latter are computed standardly by redefining the terms in the LO cross
section contributed to by the diagrams in Fig.~\ref{fig:lo} by absorbing
the self energy part of the diagram into a re-definition of the fermion
propagator (we neglect thermal effects for the heavy scalar). We have
\cite{Czarnecki:2011mr, Beneke:2014gla} therefore three sources of
thermal correction to the fermion propagator:
\begin{align}
S^{NLO}(p) & = Z_2^T \frac{\sum u^s_T(p) \overline{u}^s_T(p)} {p^2 - m_f^2 -
\Delta m_T^2}~,
\end{align}
with the fermion spin sum being thermally corrected to NLO as
\begin{align}
\sum_s u^{s}_T(p) \overline{u}^{s}_T(p) & = \slashed{p}(1-\hat{C}_B)
- \slashed{K}_B - \slashed{K}_F + m_f(1 - 2\hat{C}_B + 2 \hat{C}_F)~,
\label{eq:uubar}
\end{align}
the (finite) wave function renormalisation at temperature $T$ being
given by
\begin{align}
Z_2^T = (1+ \hat{C}_B - \hat{C}_F - 2 m_f^2 (\hat{C}'_B -\hat{C}'_F))~,
\label{eq:Z2}
\end{align}
while the thermal mass correction is given by
\begin{align}
\Delta m_T^2 = \delta m_B^2 + \delta m_F^2 + 2 m_f^2 \hat{C}_F~,
\label{eq:dm2}
\end{align}
where the circumflex indicates that the various terms are to be computed
on-shell at $p^2 = m_f^2$, and are given by
\begin{align}
C_B & = \frac{2e^2}{(2\pi)^3} \int {\rm d}^4 k \frac{\delta(k^2) n_B(\vert
k^0\vert)}{(p+k)^2 - m_f^2}~, \nonumber \\
K_B^\mu & = \frac{2e^2}{(2\pi)^3} \int {\rm d}^4 k
\frac{\delta(k^2) n_B(\vert k^0\vert) k^\mu}{(p+k)^2 - m_f^2}~,
\nonumber \\
\delta m_B^2 & = \frac{2e^2}{(2\pi)^3} \int {\rm d}^4 k
\delta(k^2) n_B(\vert k^0\vert)~,
\label{eq:B}
\end{align}
for the thermal photon contributions, and
\begin{align}
C_F & = \frac{2e^2}{(2\pi)^3} \int {\rm d}^4 t
\frac{\delta(t^2 -m_f^2) n_F(\vert t^0\vert)}{(p-t)^2}~, \nonumber \\
K_F^\mu & = -\frac{2e^2}{(2\pi)^3} \int {\rm d}^4 t
\frac{\delta(t^2 -m_f^2) n_F(\vert t^0\vert) t^\mu}{(p-t)^2}~, \nonumber \\
\delta m_F^2 & = \frac{2e^2}{(2\pi)^3} \int {\rm d}^4 t
\delta(t^2 -m_f^2) n_F(\vert t^0\vert)~,
\label{eq:F}
\end{align}
for the thermal fermion contribution\footnote{We differ in the
definition of $K_F$ from that in Ref.~\cite{Beneke:2014gla} by a sign.}.
Here the terms are expanded around $p^2 = m_f^2$ as
\begin{align}
C_{B,F} & = \hat{C}_{B,F} + (p^2 - m_f^2) \hat{C}'_{B,F}~.
\end{align}
The contributions to the thermal part of the virtual NLO annihilation
cross section from the spinor spin sum, the wave function renormalisation
and the mass corrections in Eqs.~\ref{eq:uubar}, \ref{eq:Z2} and
\ref{eq:dm2} (in effect, the NLO thermal corrections arising
from the contributions of the fermion self-energy diagram shown
in Fig.~\ref{fig:nlo_self}, and an analogous contribution from the
anti-fermion self-energy diagram shown in the same figure) are labelled
by $(B_1, B_2, B_3)$ and $(F_1, F_2, F_3)$ for the {\em thermal photon}
and {\em thermal fermion} contributions respectively. The total
contribution is the sum of all six terms.

In order to determine the divergent $K$ photon contribution (see
Eq.~\ref{eq:KG}), the thermal $K$ photon virtual contribution from Diagram
6 (and similarly from Diagram 7) of Fig.~\ref{fig:nlo_self} is calculated
as per the definition in Eq.~\ref{eq:nlo_virtual}. A straightforward
calculation of this term and comparison with the contributions listed
above shows that this matches exactly with the $B_2$ contribution and
contains within it all the soft IR divergences. (Since the fermions
are known not to contribute to the IR divergence, $F_2$ is not just
finite but vanishes). The $B_2$ contributions have thus been used in
Table~\ref{tab:Ktilde} as the virtual $K$ thermal photon contribution in
the $(i,j) = (1,1)$ and $(3,3)$ sectors in the $t$-channel, $(4,4)$
and $(6,6)$ sectors in the $u$-channel, and $(1,4)$ and $(3,6)$
sectors in the cross $tu$ channel; see the definition of $(i,j)$ in
Eqs.~\ref{eq:sigma_nlo_real} and \ref{eq:nlo_Sp} for the real photon
cross section and the explanation of the real--virtual correspondence in
Section~\ref{sssection:corresp}. Hence the soft divergent parts are only
contained in the term $B_2$ (and constitute the $K$ photon contribution),
while the collinear divergences appear in both $B_1$ and $F_1$ and are
appropriately used in Tables~\ref{tab:coll_gamma}, \ref{tab:coll_f}
and \ref{tab:coll_fbar}. The (mass correction) contribution from $B_3$
and $F_3$ are incorporated as usual via
\begin{align}
\delta \sigma & = \frac{\partial \sigma^{LO}}{\partial
m_f^2} \Delta m^2 ~, \nonumber \\
 & \equiv B_3 + F_3~.
\end{align}
Since $\sigma^{LO}$ is analytic in $m_f^2$ as seen from
Eq.~\ref{eq:sigma_LO}, both $B_3$ and $F_3$ are finite, with neither soft
nor collinear divergences.

\newpage
\printbibliography

\end{document}